\documentstyle[12pt,aasms4]{article}
\begin{document}
%
%
\newcommand\nht   {NH$_3$~}
\newcommand\hnda  {H92$\alpha$~}
\newcommand\kms    {\ifmmode{{\rm ~km~s}^{-1}~}\else{~km~s$^{-1}~$}\fi}
\newcommand\kmp    {~km~s$^{-1}$~pc$^{-1}$}
\newcommand\lo     {~$L_{\odot}$}
\newcommand\mo     {~$M_{\odot}$}
\newcommand\ro     {~$R_{\odot}$}
\newcommand\smpy   {~M_{\odot}~yr^{-1}}
\newcommand\degs   {$^{\circ}$}
\newcommand\pas    {{\rlap.}{"}}
\newcommand\pam    {{\rlap.}{'}}
\newcommand\ps     {{\rlap.}{^s}}
\newcommand\hdo    {H$_2$O~}
\newcommand\ra      {\rightarrow}
\newcommand\pdeg    {{\rlap.}{^{\circ}}}
\newcommand\Kkms    {K~km~s$^{-1}$}
\newcommand\Kkmspc  {K~km~s$^{-1}$~pc$^2$}
\newcommand\hii     {H{\small II~}}
\newcommand\nhtres  {NH$_3$~} 

\title{Massive stars: their environment and formation} 

\author{Guido Garay}
\affil{Departamento de Astronom\'\i a, Universidad de Chile,
Casilla 36-D, Santiago, Chile} 

\and

\author{Susana Lizano} 
\affil{Instituto de Astronom\'{\i}a, UNAM,
Apdo. Postal 70-264, 04510 M\'exico, D.F., M\'exico}

\begin{abstract}

Cloud environment is thought to play a critical role in determining the
mechanism of formation of massive stars. In this contribution we review the
physical characteristics of the environment around recently formed massive stars.
Particular emphasis is given to recent high angular resolution observations 
which have improved our knowledge of the physical conditions and kinematics of 
compact regions of ionized gas and of dense and hot 
molecular cores associated with luminous O and B stars. We will show that this 
large body of data, gathered during the last decade, has allowed significant 
progress in the understanding of the physical processes that take place during 
the formation and early evolution of massive stars.

\end{abstract}  


\section{INTRODUCTION}

Massive stars in our Galaxy are born predominantly within the dense cores of 
giant molecular clouds. This premise is strongly supported by a wealth of 
observations which show that hallmarks of newly formed massive stars, such as 
compact regions of ionized gas, are intimately associated with warm and dense 
regions of molecular gas (Churchwell, Walmsley, \& Cesaroni 1990; Cesaroni et al. 
1991; Plume, Jaffe, \& Evans 1992). It remains unclear, however, how the actual 
formation of massive stars takes place. In the current paradigm of the formation 
of low-mass stars (Shu, Adams, \& Lizano 1987; Shu et al. 1993), the central 
region of a dense core begins to condense quasi-statically through the process of 
ambipolar diffusion. As the magnetic pressure support decreases, the central 
region reaches an unstable quasi-equilibrium state in which the thermal pressure 
alone supports the core against its self gravity. This stage marks the initial 
condition for dynamical collapse. After becoming gravitationally unstable, the 
core undergoes a phase of free-fall isothermal collapse. As ensuing major 
evolutionary stages, the model predicts: (a) an accretion stage, characterized 
by the presence of a central protostar and a circumstellar disk surrounded by an 
infalling envelope of dust and gas; (b) a phase in which the protostar deposits 
linear and angular momentum, and mechanical energy into its surroundings through 
jets and molecular outflows; and finally, (c) a relatively more advanced phase 
in which the protostar settles onto the ZAMS, increasing its luminosity. 
Although this paradigm has been very successful in explaining what is 
observationally known about the formation of low-mass stars (cf. Lada 1991), 
its applicability to the formation of massive stars is arguable. The evolutionary 
time scales of high-mass stars are much shorter than for low-mass stars. Massive 
stars are expected to affect their environment very soon after a stellar core 
has formed since their Kelvin-Helmholtz time scale ($\le 10^4$ yrs for an O star) 
is short compared to all other relevant evolutionary 
time scales. They begin burning hydrogen 
and reach the main sequence before they stop accreting matter from the 
surrounding protostellar envelopes. The formation of a massive disk, and hence 
the appearance of molecular outflows and jets, in the accretion phase is thus not 
clear. In addition, in stage (c), the massive star starts to produce an 
appreciable output of UV photons and possibly develops strong winds which will 
drastically affect the physical conditions, structure, and chemistry of their 
surroundings. Since in this stage the massive star ionizes its surroundings, 
giving rise to a small region of ionized gas, this phase is usually referred as 
the ultracompact \hii region phase.  

An understanding of the physical processes that dominate during the early stages 
of formation of massive stars and their influence back on the molecular gas out 
of which they formed requires a detailed knowledge of the physical conditions of 
the environment prior to and after the formation of the star. Until recently, the 
observational evidence concerning the process of formation of massive stars was, 
however, scarce. Many questions remain unanswered, such as: What is the basic 
unit within a molecular cloud that will give birth to a massive star? Are massive 
stars surrounded by disks? Is the bipolar outflow phase seen in low-mass stars 
also present in the formation of high-mass stars? How is the environment around 
newly formed massive stars affected by their strong radiation and winds? What is 
the connection between massive stars and masers? Are the different maser species 
signposts of different evolutionary stages? Are massive stars formed by accretion 
processes or by stellar mergers? 
The difficulties in determining the physical conditions of the gas during the 
formation and early evolution of an individual massive star are not only due 
to their rapid evolution, but also to some observational disadvantages. 
Massive stars are born deeply embedded within molecular cores, hence the 
process of star formation is obscured by the dust that surrounds them and it 
can not be investigated at optical wavelengths. In addition, massive stars are 
usually born in clusters or groups hence their individual studies are usually 
afflicted by confusion, particularly because they are found at greater distances 
from the Sun than the sites of low-mass star formation. 

O and B stars emit the bulk of their radiation at wavelengths shorter than the 
Lyman continuum limit which ionize the dense molecular gas producing compact \hii 
regions. In addition, circumstellar dust surrounding the region of ionized gas 
absorbs all the stellar radiation, either directly or after being processed in the 
nebula, producing compact regions of warm dust that reemit the absorbed energy in 
the far infrared. Hence, the study of the environment around young massive stars 
would be better performed through observations of its ionized, atomic, and molecular 
constituents at infrared, millimeter, and radio wavelengths where the opacity of the 
dust and gas is considerably smaller. The advent of aperture synthesis radio 
telescopes have provided high angular resolution, sensitivity, and spectral 
resolution, and has opened up this field for investigation. A wealth of new 
observations of ultracompact \hii regions, photodissociation regions, cocoons of 
warm dust, hot molecular cores, and maser emission, are providing important data 
about the physical conditions and kinematics of the ionized and of the molecular gas 
very close to newly formed stars, and therefore partial answers to the questions 
posed above are beginning to emerge. 

In this article we review the results of recent high spatial resolution 
observations of high-mass star forming regions which have significantly contributed 
to the understanding of the physical conditions and dynamics, both of the ionized 
and molecular gas, in the immediate vicinity of recently formed massive stars. 
The review is organized as follows. In \S 2. we present recent 
observational data on the physical conditions and kinematics of compact \hii regions. 
Compact \hii regions tell us about the locations where massive stars form. They 
also provide information about the circumstances under which massive stars form. Do 
massive stars form in the densest part of molecular clouds? Do massive stars form in 
a highly clustered way? A number of problems that have been raised by these 
observations is also discussed. For instance, radio continuum surveys of our Galaxy 
have shown the presence of a considerable number of ultracompact \hii regions, with 
radii in the range from $\sim$0.02 pc to $\sim$0.2 pc (Wood \& Churchwell 1989a; 
Kurtz, Churchwell, \& Wood 1994). Assuming that the age of these compact \hii 
regions corresponds to their dynamical ages, then their small sizes would imply 
that they are very young objects. The large number of UC \hii regions predicted 
to exist in our Galaxy and their short dynamical ages poses a problem: {\sl the 
rate of massive star formation appears to be much greater than other indicators 
suggest} (Wood \& Churchwell 1989b; Churchwell 1990). In \S 3. we discuss the 
characteristics of the molecular gas around massive stars. Particular emphasis is 
placed on recent high angular resolution ammonia observations which have shown the 
presence of very dense, warm, and compact structures of molecular gas near massive 
stars (Garay, Moran \& Rodr\'\i guez 1993a; Cesaroni et al. 1994a; G\'omez, Garay, 
\& Lizano 1995). We address questions such as: Are the dense and warm ammonia clumps 
remnants of the process of collapse and fragmentation of a molecular core or rather 
the product of the interaction of the stellar winds and UV radiation from the 
luminous newly formed star with the surrounding natal molecular gas? Are they the 
basic units of massive molecular clouds that produce massive stars? In \S 4. we 
discuss results from recent observations of masers which are powerful signpost of 
active massive star formation. These masers provide information about the physical 
conditions and kinematics of the gas surrounding massive stars on scales of $10 -- 
10^3$ AU.  In \S 5. we discuss how massive stars may form. First we bring 
together the results presented in previous sections with other important 
observational results not dealt before which sets the observational constraints 
for possible models. Then we discuss the role of different physical processes in 
the formation of massive stars and finally a formation scenario is proposed. A 
summary is presented in \S 6.
The observed properties and models of the circumstellar dust cocoons associated 
with compact \hii regions are not covered in this article, but are nicely discussed 
in the thorough reviews of newly formed stars and their environments given by 
Churchwell (1990, 1991, 1993). 

\section{COMPACT \hii REGIONS}

Young massive stars emit copious Lyman continuum photons which excite 
their surroundings, giving rise to dense and small regions of ionized gas. These 
regions, called compact \hii regions, are characterized by having high emission 
measures which makes them very bright at radio wavelengths. Hence they act as 
powerful radio beacons of newly formed stars that are still embedded in their 
natal molecular clouds. In the first part of this section we summarize what 
is known about the morphologies, physical conditions, and kinematics of compact 
\hii regions determined from high angular resolution radio observations.
In the second part we describe theoretical models that have been proposed 
to explain the observed and derived properties of compact \hii regions. 

\subsection{Physical parameters} 

Most of the physical parameters of compact regions of ionized gas are determined 
from observations at radio wavelengths. By modelling the observed radio continuum 
spectrum as due to free-free emission arising from an isothermal and homogeneous 
region of ionized gas, the electron temperature and emission measure can be 
determined from the optically thick and thin portions of the spectrum, respectively 
(cf. Gordon 1988). In addition, if the geometry and the distance to the compact 
\hii region are known, its physical size and the average electron density can be 
derived (Mezger \& Henderson 1967). The electron temperature can also be determined 
from observations of radio recombination lines (cf. Brown, Lockman, \& Knapp 1978).

The regions of ionized gas around recently formed massive stars have diameters, $L$, 
between 0.005 and 0.5 pc, electron densities, $n_e$, between $2\times10^3$ and 
$3\times10^5$ cm$^{-3}$, and emission measures, $EM$, in the range from 
$2\times10^6$ to $1\times10^9$ pc cm$^{-6}$. These physical parameters are 
significantly correlated, as shown in Figure~\ref{fig-neemvsdiam} which presents 
plots of the electron density and emission measure against diameter for a large 
sample of compact \hii regions. The data were taken from the surveys of compact 
\hii regions of Wood \& Churchwell (1989a; squares) and Garay et al. (1993b; circles), 
and from the high angular resolution observations of compact \hii regions within the 
Sgr B2 and W49A massive star-forming regions of Gaume et al. (1995a; pentagons) and 
De Pree, Mehringer, \& Goss (1997; triangles), respectively. Least-squares linear 
fits (shown as continuous lines) to the trends, give $ n_e = 7.8\times10^2~ 
L^{-1.19\pm0.05}$ and $EM = 6.3\times10^5~ L^{-1.53\pm0.09}$, where $L$ is 
in pc, n$_e$ is in cm$^{-3}$, and $EM$ is in pc cm$^{-6}$. Since 
EM is proportional to $n_e^2 L$, these two relations are not independent of each 
other. In what follows we will only discuss the electron density versus size 
relationship. 

Assuming that the distribution of spectral types (and hence of the ionization 
characteristics) among the stars exciting compact \hii regions is independent of 
the initial conditions of the surrounding medium, then n$_e$ should be on average 
proportional to $L^{-3/2}$. A feasible explanation for the shallower observed 
power law index is that on average ultracompact ($L< 0.05$ pc) \hii regions 
are excited by stars with lower luminosities (hence lower number of ionizing 
photons) than those exciting compact ($0.05 < L < 0.5$ pc) \hii regions. This 
hypothesis is supported by Figure~\ref{fig-nivsdiam} which plots the rate of 
ionizing photons, $N_u$, versus diameter of all \hii regions plotted in 
Figure~\ref{fig-neemvsdiam} and of the \hii regions in the survey of Kurtz et 
al. (1994; stars). It clearly shows that ultracompact (UC) \hii regions are 
excited by less luminous stars than those exciting compact \hii regions. It could 
also be possible that some of the UC \hii regions in the surveys of Wood \& 
Churchwell (1989a) and Kurtz et al. (1994) do not contain luminous embedded stars 
but are instead externally ionized objects, corresponding to the denser 
structures within a larger, inhomogeneous \hii region that is excited by a single 
luminous star. This possibility is sustained by the fact that practically all the 
interstellar media has a clumpy structure (cf. Hartquist \& Dyson 1993). The number 
of ionizing photons required to excite the ultracompact source is, in this case, 
smaller than the total number actually emitted by the star by a factor of 
$\Omega/4\pi$, where $\Omega$ is the solid angle subtended by the UC clump from 
the star. Thus, the spectral type derived from the radio observations of the UC 
\hii region corresponds to a lower limit of the true spectral type of the external 
exciting star. An alternative explanation for the observed power law index is that 
interstellar dust within UC \hii regions absorbs an important fraction of the 
ionizing photons.
   
Radio continuum observations with high angular resolution show that compact regions 
of ionized gas exhibit a variety of morphologies (cf. Wood \& Churchwell 1989a, 
Garay et al. 1993b, Kurtz et al. 1994). From a survey of 75 compact \hii regions, 
Wood \& Churchwell (1989a) found that 20\% have cometary shapes - characterized by 
exhibiting a bright compact head and a diffuse extended tail-, 16\% show core-halo 
morphologies, 4\% exhibit shell structures, 17\% have irregular or multiple peaked 
brightness distributions, and 43\% are spherical or unresolved. Radio continuum 
maps of the prototype cometary (G34.3+0.15; Reid \& Ho 1985; Garay, Rodr\'\i guez, 
\& van Gorkom 1986), shell (G45.07+0.13; Turner \& Matthews 1984), and bipolar 
(Campbell 1984) compact \hii regions are shown in Figure~\ref{fig-hiimorph}. As 
noted by Wood \& Churchwell (1989a) and Fey et al. (1992),  the morphology of a 
source determined through radio synthesis observations not only depends on its 
intrinsic brightness distribution but on the response (synthesized beam) of the 
telescope. Thus, the classification depends on the sensitivity of the instrument to 
different size scales. In particular, objects classified as irregular or multiple 
peaked with very high angular resolution have been found to be bright substructures 
within more extended structures as observed with coarser angular resolution (cf. 
Garay et al. 1993b, Kurtz et al. 1999a). These could correspond to the externally 
ionized globules discussed above. The intrinsic morphology of the ionized gas, on 
the other hand, depends on the characteristics of both the exciting star and of the 
environment, as well as on their interaction. In section \S 2.3 we discuss 
theoretical models that have been proposed to explain the morphologies of compact 
\hii regions. 

\subsection{Kinematics}

Motions of the ionized gas within \hii regions may be shaped by the 
density structure of the ambient medium, by stellar winds, and/or by the motion of 
the exciting star with respect to the ambient medium. To determine the relative 
contribution of these processes a detailed knowledge of the velocity field across 
\hii regions is required. Information about the kinematics of compact \hii regions 
can be directly derived from observations of radio recombination lines, which 
yield the systemic velocity and the line width of the ionized gas. In this 
section we first discuss the observed characteristics of the integrated line 
profiles from compact \hii regions, which provide information about the global 
properties of the motions, and then the characteristics of the velocity field 
across compact \hii regions of different morphological types determined from 
observations with high angular resolution.

\subsubsection{Global motions} 

The spectral broadening of recombination lines from \hii regions includes 
three main components (cf. Gordon 1988): (1) thermal broadening, $\Delta v_{th}$, 
$$ \Delta v_{th} = \left(8~ln2 {{k~T_e}\over{m_H}}\right)^{1/2} = 
 21.4 \left( T_e \over 10^4 {\rm K}\right)^{1/2} {\rm km~s^{-1}}~~, \eqno(1)$$
where $T_e$ is the electron temperature, $k$ is the Boltzmann's constant,  
and $m_H$ is the mass of the 
hydrogen atom; (2) broadening due to electron impacts, $\Delta v_i$, given by
$$ \Delta v_i = 4.3 \left( n \over 100 \right)^{7.4} 
\left( 10^4~K \over T_e \right)^{0.1} 
\left( n_e \over 10^4 {\rm cm}^{-3} \right) ~ {\rm km~s^{-1}}~~, \eqno(2)$$ 
where $n$ is the principal quantum number of the recombination line; and  finally, 
(3) non-thermal broadening, $\Delta v_{n-t}$, which might be produced by blending 
of emission, within an observing beam, from gas at different flow velocities 
(microturbulence) or large scale velocity fields (macroturbulence). In general, 
the line profile will be a Voigt function, corresponding to the convolution of a 
Gaussian profile and a Lorentzian profile due to impact broadening. If pressure 
broadening is not important, which at the physical conditions of compact \hii 
regions is a good aproximation for recombination lines with n$\le$100, the observed 
line profile, $\Delta v_{obs}$, is roughly Gaussian and we may write
$$ \Delta v_{obs} = \left(\Delta v_{th}^2 + \Delta v_i^2 + 
\Delta v_{n-t}^2\right)^{1/2} ~~~. \eqno(3)$$

The observed parameters of radio recombination lines from compact \hii regions are 
distinctly different from those of larger regions. In particular, the line widths 
of compact \hii regions are considerable broader than those of diffuse \hii regions. 
This is illustrated in the upper panel of  Figure~\ref{fig-widvsd}, which plots the 
observed line width of recombination lines against diameter for three classes of 
\hii regions: compact (stars), typical (squares) and extended (triangles). 
As representative of typical and extended \hii regions we used, respectively, the 
sample of 36 objects observed by Churchwell et al. (1978) in the H109$\alpha$ line 
and the sample of 23 low density objects (mainly Sharpless regions) observed by 
Garay \& Rodr\'\i guez (1983) in the H125$\alpha$ line. The sample 
of compact \hii regions, listed in Table 1, consists of all compact (diameters 
$<$0.5 pc) regions of ionized gas observed to date with high angular resolution in 
either the H66$\alpha$ or H76$\alpha$ lines, which are the most commonly observed 
lines with present interferometric instruments. In order to show which is the 
dominant mechanism that produces the broad lines in compact \hii 
regions, plotted in the middle and lower panels of Figure~\ref{fig-widvsd} are the 
thermal and non-thermal line widths against the diameters of all \hii regions. The 
thermal and pressure broadening widths were computed from eqns.(1) and (2), 
respectively, using the electron temperatures derived from the radio recombination 
line observations and the electron densities derived from the radio continuum 
observations. The non-thermal widths were then computed from these and the observed 
line widths, using eqn.(3). Figure~\ref{fig-widvsd} clearly shows that the most 
important source of line broadening in compact \hii regions is of non-thermal 
origin, the thermal line widths being roughly the same in all samples. 

The non-thermal broadening of the lines may be either due to large scale systematic 
motions of the ionized gas or due to turbulence, namely the motions of small scale 
eddies within the compact \hii region. Possible sources of turbulence in \hii 
regions are the expansion of dense, small scale structures of gas into less dense 
surrounding medium; champagne flows; stellar winds, and magnetic turbulence (Arons 
\& Max 1975). As can be seen in the lower panel of Figure~\ref{fig-widvsd}, the 
non-thermal line width of \hii regions decreases as the size increases. 
Explanations for this trend have not yet been given. Two possible models that could 
explain this behavior are photoevaporation from massive circumstellar disks 
(Hollenbach et al. 1994; Yorke \& Welz 1996, Richling \& Yorke 1997) and mass 
loading from clumps embedded within UC \hii regions (Dyson, Williams, \& Redman 
1995; Lizano et al. 1996; see discussion in \S 2.5). In the photoevaporated disk
case, the line width close to the disk would reflect rotation plus expansion 
motions of the ionized gas whereas at larger distances it would only reflect 
expanding motions. In the photoevaporation from clumps model, the photoevaporated 
material that is feed into the flow is expected to increase the mean velocity 
dispersion of the region of ionized gas. Since the amount of incorporated 
photoevaporated gas is proportional to the number of ionizing photons, the observed 
trend could be due to the geometric dilution of the ionizing photons with distance. 
None of these models have, however, made explicit predictions regarding the 
dependence of line width with distance and further theoretical work is needed in 
this area. The photoevaporated disk explanation may find support on the high angular 
resolution recombination line observations of the complex W49~A massive star forming 
region which reveal that the \hii regions with the broader line widths exhibit 
also the larger spectral indices (De Pree et al. 1997). The broad line sources have 
spectral indices in the range between 0.3 to 1.0, consistent with 
those expected for ionized wind sources (Panagia \& Felli 1975).

\subsubsection{Motions within individual regions}

High angular resolution observations of radio recombination line emission are now 
making it possible to map the line center velocity and line width across compact 
\hii regions, and therefore allowing one to investigate the kinematics of the 
ionized gas within individual regions. Detailed observations of the physical 
conditions and kinematics of the ionized gas within compact \hii regions have been 
presented by Garay et al. (1986), Gaume \& Claussen (1990), Zijlstra et al. (1990), 
Wood \& Churchwell (1991), Garay, Lizano, \& G\'omez (1994), Afflerbach et al. 
(1994), and De Pree et al. (1994). In what follows we only discuss the observed 
characteristics of the velocity fields in compact \hii regions with bipolar 
morphologies since, as will be argued below, they provide information of an 
important evolutionary stage of massive stars.

Compact \hii regions with bipolar radio continuum morphologies have been found 
in only a handful of cases: NGC7538-IRS1 (Campbell 1984; Turner \& Matthews 1984), 
NGC6334(A) (Rodr\'\i guez, Cant\'o, \& Moran 1988), G45.48+0.13 (Garay et al. 1993b)
and W49A-A (De Pree et al. 1997). The radio continuum emission from the core of 
NGC7538-IRS1 shows a double lobe structure, with lobes separated by $\sim$ 600 AU. 
The prototype region, NGC6334(A), exhibits a bright compact central region, having 
a shell appearance, and extended symmetrical lobes of lower brightness. Campbell 
(1984) and Rodr\'\i guez et al. (1988) suggested that the bipolar morphology results 
from the confinement of the ionized gas by a flattened structure of neutral gas and 
dust. Radio recombination line observations of bipolar, or elongated, \hii regions 
have revealed that the line center velocity exhibits a steep gradient along their 
symmetry axis. The observed morphology and kinematics suggest that the ionized 
material in these sources is undergoing a high velocity bipolar outflow. The highly 
elongated \hii region K3-50A exhibits a change in the H76$\alpha$ line center 
velocity of $\sim$56 \kms over a linear distance of $\sim$0.37 pc, implying a 
velocity gradient along the major axis of $\sim$ 150 \kms pc$^{-1}$ (De Pree et al. 
1994). Similarly, the H76$\alpha$ and H92$\alpha$ line center velocity of the 
NGC6334A region exhibits a velocity gradient along the axis of symmetry of the 
lobes (De Pree et al. 1995a), suggesting that the lobes correspond to the two 
halves of a collimated outflow from a central star. An outflow velocity of $\sim$ 
30 \kms is derived. The largest mass motions of ionized gas detected so far are 
observed toward the core of the NGC7538-IRS1 compact \hii region. Gaume et al. 
(1995b) detected extremely wide H66$\alpha$ line profiles, of $\sim$180 \kms 
(FWHM), implying the presence of substantial mass motions. They suggest that the 
motions trace a stellar wind outflow and photoevaporation of nearby clumpy neutral 
material. These results are particularly relevant for the study of the formation 
process of massive stars, since they provide definitive evidence for the presence 
of collimated ionized bipolar outflows. In addition, they may be implying 
that massive disks, which could collimate the wind, are formed during the process 
of collapse of massive stars.

\subsection{Theoretical models}

This section presents an overview of the theoretical models proposed to explain
the characteristics of the compact \hii regions. We place particular 
emphasis on the model predictions regarding the morphologies and kinematics of the 
ionized gas, and their confrontation with observations. 
 
\subsubsection{Classical expansion} 
   
Due to the difference in pressure between the ionized gas and the surrounding
neutral gas, \hii regions are expected to expand in the neutral ambient medium.
The classical analysis of the expansion of \hii regions (cf. Spitzer 1978; 
Dyson \& Williams 1980) assumes that the ambient medium is homogeneous 
in density and temperature and neglects the role of stellar winds. 
The expansion in this case is characterized by two main phases. In the initial 
stage, which starts when the young massive star embedded in a molecular cloud 
begins to produce UV photons, an ionization front is formed which moves rapidly 
outward through the ambient medium. This phase of expansion comes to an end when 
the number of photoionizations within the ionized region equals the number of 
recombinations. At this point the \hii region fills a region with a radius, 
$R_{S}^{o}$ called the Str\"{o}mgren radius, given by (Str\"omgren 1939) 
$$ R_{S}^{o} =  0.032 
\left({{N_{\rm u}}\over{10^{49}~s^{-1}}}\right)^{1/3} 
\left({{10^{5}~cm^{-3}}\over{n_o}}\right)^{2/3}~~{\rm pc}, \eqno(4)$$ 
where $n_o$ is the initial density of the ionized gas (= $2n_{H_2}$, where 
$n_{H_2}$ is the molecular density of the ambient gas), and {$N_{\rm u}$} 
is the rate of ionizing photons emitted by the exciting star. 
This initial phase is rapid and short lived. The characteristic time, $t_S$, 
in which the Str\"{o}mgren radius is reached is given by  
$$ t_S = {{1}\over{n_o\alpha_{B}}}~~~,  $$
where $\alpha_{B}$ is the recombination coefficient excluding captures
to the ground level. For an O7 ZAMS star, which has an output of UV photons 
of 4$\times10^{48}$ s$^{-1}$ (Panagia 1973), born in a medium with an ambient 
density of 10$^5$ cm$^{-3}$, the Str\"omgren radius, of 0.015 pc, is reached in 
$\sim1$ year. This, however, should only be taken as a fiducial value 
since it is likely that the star turns on its output of UV photons
in a lapse of time considerably larger than this. 

Afterwards, the heated gas will expand and form a shock front that moves out 
through the neutral gas. The rate of expansion of the \hii region during this phase 
is then determined by the interaction between the ionization and shock fronts.
In this phase the radius, R$_i$, of the \hii region increases with time, t, as
(cf. Spitzer 1978) 
$$ R_i = R_{S}^{o} \left(1+ {{7~C_{\tiny II}~t}\over{4~R_{S^o}}}\right)^{4/7} ~, 
\eqno(5)$$ 
where $C_{\small II}$ is the sound speed in the ionized gas. The expansion stalls 
when the hot, but lower density, ionized gas reaches pressure equilibrium with 
the surrounding cool ambient medium. The final equilibrium radius of the \hii region, 
$R_f$, is (Dyson \& Williams 1980)
$$ R_f = \left({{2 T_e}\over{T_o}}\right)^{2/3} R_{S}^{o}~~, $$  
where $T_o$ is the temperature of the ambient gas. Using eqn.(4), 
the equilibrium radius can be written as
$$ R_f =  1.1 
\left({{N_{\rm u}}\over{10^{49}~s^{-1}}}\right)^{1/3} 
\left({{T_e}\over{10^4 K}}\right)^{2/3} 
\left({{100 \rm K}\over{T_o}}\right)^{2/3} 
\left({{10^{5}~cm^{-3}}\over{n_o}}\right)^{2/3}~~{\rm pc}, \eqno(6)$$ 
and is reached in a lapse of time, $t_{eq}$,
$$ t_{eq} \leq 7.6\times10^5 
\left({{N_{\rm u}}\over{10^{49}~s^{-1}}}\right)^{1/3} 
\left({{10^{5}~cm^{-3}}\over{n_o}}\right)^{2/3} 
\left({{T_e}\over{10^4 K}}\right)^{2/3} 
\left({{100~ \rm K}\over{T_o}}\right)^{7/6}~~{\rm yrs}. \eqno(7)$$ 
The upper limit arises because in the derivation of eqn.(5) the pressure 
of the ambient gas has been neglected.

\subsubsection{Stellar winds} 

The expansion due to the gradient in pressure is not, however, the only way in which 
the interstellar medium around a young massive star can be set in motion. Luminous, 
massive stars are known to possess strong stellar winds, and it is possible that the 
interstellar gas motions are dominated by the wind energy and momentum. Thus, winds 
may play an important role in the dynamical evolution of compact regions of ionized 
gas. In fact, stellar winds (Castor, McCray, \& Weaver 1975; Shull 1980) have been 
proposed as the main mechanism for producing the shell morphologies observed in UC 
\hii regions. Another mechanism that could produce a shell structure is radiation 
pressure on dust grains (Kahn 1974), but this has been shown to be less important 
than stellar winds (Turner \& Matthews 1984).

Massive stars have powerful winds which are expected to cause a considerable
impact on their environment as they deposit momentum and mechanical energy into the 
interstellar medium (Van Buren 1985). It is not known, however, when the wind 
phenomenon begins in massive stars. Probably well before a stellar wind is set into 
motion, bipolar outflows produced in the formation process will have a major input 
of mechanical energy and momentum. Assuming that the dynamical effects of the wind 
are important from the very beginning of the compact \hii region phase then the 
evolution proceeds roughly as follows. The interaction of the stellar wind with the 
interstellar medium produces a dense shell of circumstellar gas that expands away 
from the star. This shell is exposed to the UV radiation from the recently formed 
star, thus it will be either totally or partially ionized. The evolution of the \hii 
region is thus closely tied to the evolution of the circumstellar shell. When the 
shell is driven by the pressure of the hot bubble of shocked stellar wind, the 
radius of the shell, R$_{sh}$, increases with time as (Castor et al. 1975) 
$$ R_{sh} = 0.042 
\left({{L_w}\over{10^{36}~ergs~s^{-1}}}\right)^{1/5} 
\left({{n_o}\over{10^5~cm^{-3}}}\right)^{-1/5} 
\left({{t}\over{10^3~years}}\right)^{3/5} ~{\rm pc},  \eqno(8)$$
where $L_w$ is the wind mechanical luminosity. 
The velocity of expansion of the shell, $V_{\rm sh}$, is then 
$$ V_{\rm sh} = 24.7 
\left({{L_w}\over{10^{36}~ergs~s^{-1}}}\right)^{1/5} 
\left({{n_o}\over{10^5~cm^{-3}}}\right)^{-1/5} 
\left({{t}\over{10^3~years}}\right)^{-2/5} ~{\rm \kms}. \eqno(9)$$

UC \hii regions with shell or ring structures, determined from observations 
with high angular resolution, have been reported by Turner \& Matthews (1984)
and by Garay et al. (1986). Turner \& Matthews (1984) concluded that while the 
observed shell-like morphologies can be well explained as due to the action of 
stellar winds, the radiation pressure model faces several difficulties. 
In particular the morphology and velocity structure of the G45.07+0.13 compact 
\hii region, observed with 0\pas4 angular resolution, is well modeled by a ring 
of ionized gas expanding with a velocity of $\sim$10 \kms driven by the wind of 
a star with a mechanical luminosity of $\sim1\times10^{35}$ ergs s$^{-1}$
(Garay et al. 1986). Acord, Churchwell, \& Wood (1998) and Kawamura \& Masson 
(1998) measured the angular expansion rate of the shell-like UC \hii regions 
G5.89-0.39 and W3(OH), respectively, deriving dynamical ages of $\sim 600$ and 
$\sim 2300$ years. In the case of G5.89-0.39, assuming that the wind mechanical 
luminosity of the O6 central exciting star is $\sim 3\times10^{36}$ ergs~s$^{-1}$
(cf. Van Buren 1985) and that the density of the ambient medium is in 
the range $10^7-10^8$~cm$^{-3}$ (Harvey et al. 1994), then the observed shell 
radius and expansion velocity can be explained by the simple stellar wind model 
if the nebula is very young, $\sim 5\times10^2$ years, consistent with the 
derived dynamical age. In the case of the W3(OH) object, Kawamura \& Masson (1998) 
find that there is sufficient ram pressure from the stellar wind to sustain 
the shell structure. All these observations attest that winds 
are present from a very early stage in the evolution of newly formed stars. 

The question that arises is: Under what conditions is the wind dynamically more 
important than the classical expansion due to the difference in pressure between 
the ionized gas and ambient medium? This has been investigated by Shull (1980) 
and by Garay et al. (1994) who studied the conditions under which a 
stellar wind in a medium with density gradients remains confined inside the flow 
driven by the difference in pressure alone. Shull (1980) concluded that the wind 
is more important when
$$ \left({{L_w}\over{10^{36}~ergs~s^{-1}}}\right) > 0.33 
\left({{N_{\rm u}}\over{10^{49}~s^{-1}}}\right)^{2/3} 
\left({{n_0}\over{10^5 cm^{-3}}}\right)^{-1/3} ~~. $$
The observational data that could permit one to test this theoretical prediction are 
not yet available, mainly due to the difficulty of deriving wind 
mechanical luminosities.

The classical expansion and the stellar wind models predict that \hii regions should 
have spherically symmetric morphologies, and hence they do not adequately explain 
most of the observed morphologies. In what follows we will discuss 
models that have been proposed to explain non-spherical morphological types.  

\subsubsection{Champagne flows} 

Champagne models assume that the medium in which a massive star is born is not 
uniform but has strong density gradients, which gives rise to an \hii region that 
expands supersonically away from the high density region in a so called champagne 
flow (Tenorio-Tagle 1979, Bodenheimer, Tenorio-Tagle, \& Yorke 1979). As for the 
classical model, the 
evolution of the region of ionized gas is characterized by two main phases. In the 
initial stage, the ionization front rushes rapidly into the ambient medium. If the 
density falls off faster than r$^{-3/2}$ the ionization front in the direction of 
decreasing density is not trapped and the whole zone facing the low density region 
becomes ionized (Franco, Tenorio-Tagle, \& Bodenheimer 1990). The large pressure 
gradient left behind by the ionization front induces the formation of a strong 
shock which moves supersonically into the ionized low density medium. This marks 
the beginning of the second stage of evolution, in which the ionized gas begins to 
stream away toward the direction of decreasing density.

Theoretical radio continuum maps of \hii regions in the champagne phase, namely 
when an expanding \hii region within a molecular cloud reaches the cloud's
edge, have been presented by Yorke, Tenorio-Tagle, \& Bodenheimer (1983). The
simulations show that during this phase the resulting configuration is a blister 
type \hii region, which is ionization bounded on the high density side and density 
bounded on the side of outward champagne flow. The morphology of the ionized gas 
is roughly shaped like an opened fan. Numerical calculations of the velocity 
structure of the ionized gas during the champagne phase show that the ionized 
material is accelerated in the direction away from the molecular cloud to a 
relatively high velocity, up to several times its sound speed (Yorke, Tenorio-Tagle, 
\& Bodenheimer 1984). Although the velocity increases with increasing distance 
from the cloud, and can attain values in excess of 30 \kms, the average velocity of 
the ionized gas, integrated over the whole \hii region, is shifted by a small 
amount ($\le$5 \kms) with respect to the velocity of the molecular cloud.  

A clear demonstration of the presence of champagne flows associated with compact 
\hii regions has been provided by high angular resolution observations of radio 
recombination line emission, which have permitted determination of the velocity 
structure 
across the compact sources. The two fan-shaped G32.80+0.19 B and G61.48+0.09 B1 
compact \hii regions exhibit striking gradients in the velocity of the ionized gas 
running along of the symmetry axis of the cometary-like structures, the velocity 
increasing smoothly from the head leading edge to the tail, by $\sim$8 and 12 \kms, 
respectively (Garay et al. 1994). The observed velocity fields and morphologies, 
which are in good agreement with those predicted by the champagne model, 
combined with the similar velocities of the ionized gas at the head position and 
of the molecular gas, imply that these regions are undergoing champagne 
flows. Toward the compact \hii region W3(OH), Keto et al. (1995) detected in the 
bright region of ionized gas a velocity gradient of 12 \kms, as well as weak radio 
continuum emission aligned with the velocity gradient and extending away from the 
high density gas immediately around the newly formed star. They interpret these 
observations as indicating that the W3(OH) \hii region is undergoing a supersonic 
champagne flow (see also Sams, Moran \& Reid 1996). The shift in the recombination 
line velocity with principal quantum number observed toward this source (Berulis \& 
Ershow 1983; Welch \& Marr 1987), which is not explained within the classical model, 
is succesfully modeled in terms of blending of emission from gas at different 
velocities and densities within the champagne flow model.  

\subsubsection{Bow shocks} 

An alternative hypothesis proposed to explain cometary morphologies is that 
cometary \hii regions correspond to bow shocks supported by the stellar wind of 
an ionizing star that is moving supersonically through the ambient molecular cloud
(Van Buren et al. 1990; MacLow et al. 1991). The bow shock model is able to 
explain fine observational aspects in the morphology of certain cometary \hii 
regions, such as limb brightening and/or that the emission pinchs back down to 
the symmetry axis, which are not easily explained within the champagne models.
The characteristic size of a cometary \hii region in the bow shock model 
is provided by the distance in front of the star where the terminal wind 
shock occurs. This distance, $l_{bs}$, where the momentum flux in the wind equals 
the ram pressure of the ambient medium is (Van Buren et al.~1990) 
$$ l_{bs} = 0.015 
\left({{\dot{M}_*}\over{10^{-6}~\smpy}}\right)^{1/2} 
\left({{v_w}\over{10^3~\kms}}\right)^{1/2} $$ 
$$~~~~~~~~~~\times  \left({{n_o}\over{10^5~cm^{-3}}}\right)^{-1/2} 
\left({{v_*}\over{10\kms}}\right)^{-1}
\left({{\mu_H}\over{1.4}}\right)^{-1/2} ~{\rm pc}, \eqno(10)$$
where $\dot{M}_*$ is the stellar wind mass-loss rate, $v_w$ is the wind terminal 
velocity, $v_*$ is the velocity of the star relative to the molecular cloud, 
and $\mu_{\rm H}$ is the mean mass per hydrogen nucleus. 

Detailed models of the expected velocity structure of the ionized gas for the 
bow shock hypothesis have been presented by Van Buren \& MacLow (1992). 
As discussed previously, both the champagne and bow shock models reproduce well 
the observed morphology of cometary \hii regions. The main differences between 
these two models concern the predictions about the kinematical properties of the 
ionized gas and lifetimes. The bow shock model predicts that the velocity gradient 
should be steeper in the head than in the tail, whereas in the champagne model the 
largest gradients are expected in the tail, where the gas is accelerated out the 
nozzle.  In addition, contrary to the champagne flow model, 
the bow shock model predicts that the line widths should 
be broader along the leading edge of the ionization front than they are behind.  
Finally, the champagne flow model predicts that the velocity of the ionized gas near 
the coma of the cometary structure should be at rest with respect to the molecular 
gas velocity while the bow shock model predicts it should be moving with the 
velocity of the star. Cometary \hii regions for which the observations 
seem to be best explained by a bow shock model are G29.96-0.02 (Wood \& Churchwell 
1991; Van Buren \& Mac Low 1992; Afflerbach et al. 1994), and G13.87+0.28 
(Garay et al. 1994). Fey et al. (1995) and Lumsden \& Hoare (1996) 
have however questioned the bow shock interpretation for G29.96-0.02,
arguing in favor of a champagne flow. These contradictory results 
reflect the difficulty in observationally discriminating between the
bow shock and champagne models.

Other models that have been proposed to explain cometary regions include 
supersonic Str\"{o}mgren regions (Raga 1986), distorted magnetic
\hii regions (Gaume \& Mutel 1987), and distorted stellar wind bubbles (Mac Low 
\& McCray 1988).

\subsection{Clustering and clumpiness} 

Many compact \hii regions are often found in groups and/or in the vicinity of
larger, diffuse \hii regions (Habing \& Israel 1979; Wood \& Churchwell 1989a; 
Garay et al. 1993b; Tieftrunk et al. 1997). This is illustrated in 
Figure~\ref{fig-hiigroup} which presents a 4.9 GHz image of the W3 Main star 
forming region showing the presence of a cluster of internally ionized compact 
\hii regions, and therefore of the presence of a recently formed association of 
O and B stars (Colley 1980; Tieftrunk et al. 1997). There are more than 10 \hii 
regions, with diameters ranging from 0.01 to 0.7 pc and excited by stars with 
spectral types ranging from B1 to O6, within a region of $\sim$2 pc in diameter. 
To investigate quantitatively the gregarious nature of compact \hii regions, Garay 
et al. (1993b) observed with arcsec angular resolution the radio continuum emission 
toward 16 highly luminous IRAS point sources known to be associated with compact 
\hii regions detected with single dish telescopes. They found that a large 
fraction of these sources, particularly the most luminous ones, exhibit complex 
radio morphologies which could be decomposed into multiple components, strongly 
supporting the premise that massive stars tend to be born in rich groups or 
clusters. They conclude that as many as 50\% of the IRAS point sources associated 
with unresolved (by single dish instruments) radio continuum sources, might be 
excited by a cluster of young O and B stars rather than by a single star. 

The complex morphology of regions of ionized gas may not always, however, 
signal the presence of a cluster of O and B stars. The multiple peaked or 
irregular morphologies of compact \hii regions may well reflect the presence of 
density inhomogeneities within the ambient medium. Molecular clouds are known to 
be clumpy and highly inhomogeneous. In addition, the presence of clumps near 
recently formed stars might be further enhanced by the fragmentation process 
that is likely to take place during the gravitational collapse. The important 
question whether the multiple compact components mark the location of young O stars 
or just the position of clumps of gas, was first raised by Dreher et al. (1984). It 
is not clear yet what fraction of the ultracompact \hii regions associated with 
more diffuse \hii regions truly corresponds to regions excited by an embedded star 
and which corresponds to dense clumps within an inhomogeneous region that is being 
externally excited by a single luminous star which ionizes both the compact region 
and the larger diffuse region. 

\subsection{Ages}

Regions of ionized gas born in a medium of constant density are expected to 
expand into the ambient neutral gas at the sound speed of the ionized gas, 
until they reach pressure equilibrium with the ambient medium. Thus, if \hii regions 
are smaller than $R_f$ (see eqn.[6]), then their sizes would indicate their ages. 
For instance, eqn.(5) shows that a region of ionized gas excited by an O7 star, 
born in a medium with a molecular ambient density of $10^5$ cm$^{-3}$, would have 
expanded to a radius of 0.1 pc in a lapse of time of $\sim2\times10^4$ years. The 
small sizes of the UC \hii regions (R$\leq$0.05 pc) would then imply that they are 
very young objects, with lifetimes $\le$ $5\times10^3$ years. 

Using the IRAS Point Source Catalog (PSC) and a color criterion that they propose 
selects UC \hii regions, Wood \& Churchwell (1989b) estimated that there are 
potentially $\sim$1650 UC \hii regions within the disk of our Galaxy. Assuming that 
these objects have physical parameters similar to those actually derived 
for a small number of UC \hii regions (namely, radii of $\sim$0.05 pc, number of 
ionizing photons of $\sim4\times10^{48}$~s$^{-1}$; Wood \& Churchwell 1989a)
and that they are born in a medium with an ambient density of $\sim10^5$ cm$^{-3}$,
then their dynamical ages are typically $\sim5\times10^3$ yr. The implied rate of 
massive star formation, $\psi\sim$ 0.3 O stars yr$^{-1}$, is considerably 
larger than that estimated from other means. For instance, G\"usten \& Mezger (1983; 
see also Downes 1987) estimate that the star formation rate of OB stars ($10<M<60$ 
\mo) is $\sim$0.82 \mo~yr$^{-1}$, which using an IMF-weighted typical mass of 
23\mo~ it implies that $\psi\sim4\times10^{-2}$ O stars yr$^{-1}$, about 10 times 
smaller than that derived under the above hypothesis. Note, however, that 
Codella, Felli, \& Natale (1994a) found that 65\% of a large ($\sim 450$) sample 
of diffuse \hii regions they studied are associated with IRAS point sources 
and satisfy the Wood \& Churchwell color criteria. This criterion, therefore, 
selects compact as well as more diffuse \hii regions, which may alleviate the 
problem. 

Why do we see so many UC \hii regions? Or phrasing it differently, Why the ages of 
UC \hii regions are not consistent with their short dynamical ages? Wood \& 
Churchwell concluded that the expansion of UC \hii regions is inhibited by some 
mechanism, so that their small sizes do not necessarily indicate that they are 
extremely young. Several suggestions have been made to explain the lifetime 
paradox.  Van Buren et al.~(1990) proposed that most UC \hii regions are excited 
by O stars that are in motion relative to the molecular cloud and that have formed 
a bow shock supported by its stellar wind (see  \S 2.3.4). Since bow shocks are static 
configurations, neither expanding nor contracting, UC \hii regions would {\sl live} 
as long as the moving star remains embedded within the molecular cloud. Hollenbach 
et al. (1994) proposed that newly formed OB stars are surrounded by a massive 
primordial disk which is photoevaporated by the UV photons from the star.  
The dense gas that is ionized close to the star, and that flows away from it, 
would then give rise to the observed UC \hii region. The reservoir of dense gas 
within the disks may last for a million years or more, depending on the mass of 
the disk. Thus, UC \hii regions could live much longer ($\sim3\times10^5$ years) than 
what their dynamical ages indicates because they are constantly being replenished 
by a dense circumstellar reservoir. Whereas the formation of accretion disks around 
massive young OB stars is expected given that molecular clouds have non-zero angular 
momentum, direct observational evidence for its presence is scarce, although 
beginning to emerge (see \S 5.1.3). 

Several authors (Dyson 1994; Dyson et al. 1995; Lizano \& Cant\'o 
1995; Lizano et al. 1996) have shown that the identification of the dynamical time 
scale with the age of compact \hii regions is not suitable if the primordial ambient 
medium is clumpy. The presence of clumps near recently formed stars is 
expected due to the fragmentation process that takes place during the gravitational 
collapse. Dyson et al. (1995) investigated the interaction of the wind of a young 
massive star with the clumpy molecular gas and showed that mass loading due to the 
hydrodynamical ablation of clumps could result in long lived UC \hii regions.
Lizano et al. (1996) showed that the mass injection from photoevaporated globules 
into a stellar wind causes trapping of the ionization front resulting in compact 
(0.1 pc) long-lived ($\sim10^5$ yrs) \hii regions. Whereas both processes occur, 
Arthur \& Lizano (1997) showed that photoevaporation dominates over ablation for 
self-gravitating clumps immersed in \hii regions.  

Alternatively, De Pree, Rodr\'\i guez, \& Goss (1995b) (see also Garc\'\i a-Segura 
\& Franco 1996; Xie et al. 1996) suggest that the paradox finds a simple explanation 
if the physical conditions of the ambient medium are much denser and warmer than 
previously believed resulting in very small equilibrium radii, of the order of the 
size of UC \hii regions. This hypothesis attains support on recent 
observations which show that the temperatures and densities of the molecular gas 
around UC \hii regions are typically $\sim$100 K and $\sim1\times10^7$ cm$^{-3}$, 
respectively (see \S 3.3). Under these conditions, the equilibrium radius of a 
region of ionized gas excited by an O9 star ($N_u=1.2\times10^{48}$ s$^{-1}$) is 
just $\sim$0.016 pc and the time it takes to reach pressure equilibrium is 
$\sim 1\times10^4$ years. Thus most UC \hii regions could be objects that have 
already reached the equilibrium radius and hence be much older than what the 
dynamical age indicates. This possibility is further strengthened by the fact that 
a large fraction of UC \hii regions are excited by stars emitting an output rate of 
ionizing photons smaller than that of an O9 star (see \S 2.1), and therefore
should have smaller equilibrium radii. 

\vfill\eject
  
\section{HOT MOLECULAR CORES}

A wealth of observations have established that the presence of dense 
molecular gas toward regions of newly formed massive stars is quite common. 
For instance, a survey of emission in the CS(7$\rightarrow$6) transition towards 
a large sample of star forming regions associated with water masers show detection 
in $\sim$60\% of the cases (Plume et al. 1992). Since this transition 
has a critical density of $\sim2\times10^7$ cm$^{-3}$, the detection of emission 
indicates the existence of very dense ($n(H_2)> 10^6$ cm$^{-3}$) gas within 
molecular cores. In addition to the high densities, observations of highly excited 
inversion transition lines of ammonia have shown that the presence of hot molecular 
gas, with temperatures in the range from 100 to 250 K, is also common (Mauersberger 
et al. 1986; Cesaroni, Walmsley, \& Churchwell 1992). These observations, which 
were carried out using single dish instruments, lack adequate angular 
resolution to determine the location of the hot and/or dense molecular emission with 
respect to the compact \hii regions, and to resolve its spatial structure and 
kinematics. 

In this section we discuss the results of recent studies of dense and hot 
molecular emission, made with high angular resolution ($\le 5$''), which have 
yielded information about their structure, physical properties, and kinematics (eg., 
Heaton, Little, \& Bishop 1989; Garay \& Rodr\'\i guez 1990; Garay et al. 1993a; 
Cesaroni et al. 1994a,b; G\'omez et al. 1995; Garay et al. 1998; Cesaroni et al. 
1998). These observations show that the hot (T$_{\rm K} >$ 50 K) and usually dense 
($>10^5$ cm$^{-3}$) molecular gas arises from small ($<0.1$ pc) structures with 
masses in the range  $10^2 - 3\times10^2$ \mo. We will refer to these structures 
as hot molecular cores. We note that although some authors use this term to imply 
the presence of an embedded object, we use it here just to refer to hot and 
compact structures of molecular gas regardless of whether they are internally or 
externally illuminated objects.  Particular emphasis will be given to the physical 
and dynamical association between the hot cores and compact \hii regions, and to 
their evolutionary stage. 

The high temperatures of hot cores increases evaporation of icy grain mantles, 
making the chemical processes in the vicinity of UC \hii regions different from 
that in cold dark clouds. In particular, the gas phase is expected to be enriched 
with mantle constituents such as NH$_3$, H$_2$O, and CH$_3$OH. A comprehensive 
review of the 
observations and models of the chemistry in hot cores has been recently presented by 
van Dishoeck \& Blake (1998). In dense, warm molecular cores the abundance of 
ammonia is considerably enhanced with respect to that in cold dark clouds (cf. 
Walmsley 1989, 1990). Thus, NH$_3$ lines, particularly those with high excitation 
energy levels, have proved to be excellent tracers of the dense molecular gas 
around newly formed stars. Accordingly most of the high angular resolution 
observations of compact cores have been made in lines of ammonia. 

\subsection{The habitat: massive cores}

Hot and dense molecular cores are not isolated entities but are compact 
features within larger, less dense, structures of molecular gas. Molecular clouds 
are known to be highly inhomogeneous clumpy entities exhibiting a hierarchy 
of structures with densities spanning several orders of magnitude (e.g. Wilson \& 
Walmsley 1989; Blitz 1993). In this subsection we shortly summarize the 
characteristics and physical conditions of the molecular environment that 
surrounds hot cores and compact regions of ionized gas derived from 
observations made with single dish instruments (HPBW of typically 40") which 
effectively provide the characteristics of the molecular emission averaged over 
spatial scales of typically $\sim$1 pc. We assume that these data are 
representative of the conditions of massive cores (Caselli \& Myers 1995). 
Massive cores are themselves not isolated entities but are found within Giant 
Molecular Clouds (GMCs). They fill only a small fraction of the total GMC area 
and are concentrated in regions of active star formation. About 20\% of the total 
gas in GMCs is found in the form of massive cores (cf. Tieftrunk et al. 1998a).

The derived parameters of massive cores depend on the molecular line used 
to trace them. Churchwell et al. (1990; hereafter CWC90) undertook a survey of 
ammonia emission in the (1,1) and (2,2) inversion transitions toward a large sample 
of compact \hii regions detecting emission toward 70\% of them, implying that young 
regions of ionized gas are still embedded within massive cores, and deriving ammonia 
column densities of typically $\sim 10^{15}$ cm$^{-2}$ and gas kinetic temperatures 
of typically 28 K. The line widths of the ammonia emission range between 1.5 and 4 
\kms, with an average value of $\sim 3.1$ \kms. Since the thermal part of the line 
width amounts to only $\sim0.3$ \kms (for T$_K \sim 28$ K), 
the line widths are largely dominated by non-thermal motions. 
Cesaroni et al. (1992) observed sixteen massive cores associated 
with compact \hii regions in the (4,4) and (5,5) transitions of \nht, which are 
thought to trace warmer and denser gas than the (1,1) and (2,2) transitions, 
deriving rotational temperatures of typically $\sim$ 46 K, ammonia column densities 
of typically $4\times10^{16}$ cm$^{-2}$, and molecular densities of $\sim10^6$ 
cm$^{-3}$. From observations of CS emission around compact regions of ionized gas, 
Cesaroni et al. (1991) found massive cores with densities of $\sim10^6$ cm$^{-3}$, 
sizes of $\sim0.4$ pc, and masses of $\sim 2\times10^3$ \mo. In summary, the low 
angular resolution observations show that massive cores associated with regions of 
recent massive star formation have linear sizes between 0.3 - 1.0 pc, kinetic 
temperatures in the range $30 - 50$ K, molecular densities in the range 
$2\times10^4-3\times10^6$ cm$^{-3}$, and masses between $10^3 - 3\times 10^4$ \mo. 
Caselli \& Myers (1995) found that massive cores associated with embedded young 
stellar objects have physical properties almost identical to neighboring massive 
starless cores. Thus, these physical conditions are probably representative of 
massive cores in a phase prior to the formation of an OB cluster. 

Since massive cores contain hot and dense structures it is clear that the 
molecular gas within them is neither homogeneous nor isothermal. To determine how 
the physical and chemical conditions of the gas within a single massive core varies 
as a function of the distance from its center is a laborious task. It requires 
mapping in a variety of molecular transitions and with different angular 
resolutions. One of the most extensively scrutinized massive star forming region 
is G34.3+0.15, which has been observed with angular resolutions 
ranging from $\sim2'$ (single-dish) to $\sim1''$ (aperture synthesis) and in 
several molecular transitions: at low angular resolutions (40$''$ to 130$''$) in 
lines of NH$_3$ (Heaton et al. 1985), at intermediate angular resolutions 
($20''-40''$) in several rotational transitions of CO, HCO$^+$ and isotopes 
(Matthews et al. 1987; Heaton et al. 1993), and interferometrically in several 
lines of NH$_3$ (Andersson \& Garay 1986; Heaton et al. 1989; Garay \& Rodr\'\i 
guez 1990) and HCO$^+$, H$^{13}$CN, HC$^{15}$N, and SO (Carral \& Welch 1992). The 
observed morphology of the molecular emission at different spatial scales and in 
different chemical probes is illustrated in Figure~\ref{fig-g34morph}, which shows 
maps of: a) NH$_3$ emission from observations with angular resolution of $2\pam2$ 
(Heaton et al. 1985); b) HCO$^+$ emission with $6''$ angular resolution (Carral 
\& Welch 1992); and c) NH$_3$ emission with $1''$ angular resolution (Heaton et 
al. 1989). At the largest linear scales, the ammonia observations trace an 
extended (diameter $\sim 3.7$ pc), cool (T $\sim$9 K), low density (n(H$_2)\sim 
5\times10^3$ cm$^{-3}$) {\sl halo} surrounding a denser ($\sim 4\times10^4$ 
cm$^{-3}$) {\sl core} having a diameter of $\sim$1.7 pc (Heaton et al. 1985). The 
HCO$^+$ emission traces a warm (T$\ge 25$K) structure of about 0.9 pc in diameter. 
The NH$_3$ and SO interferometric observations made with angular resolutions of 
$\sim5''$ reveal the presence of a warm ($\sim60$K), dense (6$\times10^5$ cm$^{-3}$), 
and small ($0.21\times0.05$ pc) substructure within the core, referred to as the 
{\sl compact core}. At higher resolution ($\sim1''$), Heaton et al. (1989) detected 
an {\sl ultracompact core} which has a molecular hydrogen density of $\sim4\times10^7$ 
cm$^{-3}$ and a linear size of $0.05\times0.02$ pc. 
The hot ammonia emission from the ultra compact dense 
core originates in a small region lying at the eastern edge and approximately 
delineating the leading edge of the ultra compact cometary \hii region. This 
morphology suggests that the emission from the ultra compact core arises from 
molecular gas that has been compressed either by the expansion of the \hii region, 
by stellar winds, or by the relative motions of an O star within the molecular cloud. 

Using the data available on ammonia emission, which covers linear scales ranging 
from a few hundredths to a few parsecs, Garay \& Rodr\'\i guez (1990) determined
that the density and temperature of the molecular gas toward G34.3+0.15 decrease with 
distance from the peak continuum emission, $r$, as $r^{-1.7\pm0.4}$ and 
r$^{-0.6\pm0.1}$, respectively. Using HCO$^+$ data, Heaton et al. (1993) derived 
that the molecular density is constant out to a radius of 0.1 pc and decreases 
outwards with radius as $r^{-2.1}$. Given the uncertanties in both determinations, 
such as in the adopted molecular abundances, the agreement is good. 
The density dependence with radius determined by Garay \& Rodr\'\i guez (1990) is 
similar to that implied by Caselli \& Myers (1995) for a large sample of massive 
dense cores, in 
the context of an equilibrium model of a dense core with thermal and non-thermal 
motions. They find that the density profile goes like $r^{-1.6}$ in massive dense 
cores and $r^{-1.1}$ in low-mass dense cores. Massive cores are, however, likely 
to be highly fragmented into filaments and dense clumps, thus the physical 
conditions derived assuming smooth density and temperature profiles should be 
taken only as representative of the bulk conditions of the clumpy emitting region. 
Whether the derived physical conditions of the G34.3+0.15 massive core are 
representative of that of the molecular gas in most of the massive cores forming 
massive stars has yet to be observationally established. In any case, the above 
observations clearly illustrate that hot cores are not isolated entities 
but are rather dense structures embedded in cooler, lower density, and more extended 
structures which dominate the emission detected in low angular resolution studies. 

\subsection{Association with compact \hii regions}

The study of the spatial distribution of the highly excited molecular emission 
relative to the thermal emission of the ionized gas on scales of $<$0.1 pc has 
began only recently, via high angular resolution observations of NH$_3$ made with 
the VLA (Heaton et al. 1989; Garay et al. 1993a; Cesaroni et al. 1994a; G\'omez et 
al. 1995; Garay et al. 1998; Cesaroni et al. 1998), and CH$_3$CN made with the 
IRAM interferometer (Cesaroni et al 1994b; Olmi et al. 1996). These observations 
show that the highly excited emission arises from compact, hot molecular structures 
that are invariably located near and in most cases intimately associated with 
compact \hii regions. Hot cores are thus found in the immediate vicinity of newly 
formed massive stars. In some cases the observed line profiles appear in 
emission off to the side of the \hii region and in absorption in front of it (cf. 
Keto, Ho, \& Haschick 1987a; Garay \& Rodr\'\i guez 1990; Gaume \& Claussen 1990). 
In other cases the lines are only seen in absorption (W3(OH), Reid, Myers, \& 
Bieging 1987; NGC 7538, Henkel, Wilson, \& Johnston 1984), implying that the bulk 
of the dense molecular gas is located in front of the compact \hii region.
The intimate association is illustrated in Figure~\ref{fig-nh3cores} which shows 
maps of the hot ammonia and radio continuum emission toward the G10.47+0.03 
and G29.96-0.02 (Cesaroni et al. 1998) and G61.48+0.09 (G\'omez et al. 
1995) regions of young massive star formation. 
Not all compact \hii regions are associated with dense and warm ammonia 
clumps, however. For instance, observations of the more than a dozen 
compact \hii regions in the Sgr B2 region show that only two of them are 
associated with dense and hot ammonia clumps (Vogel, Genzel, \& Palmer 1987; 
Gaume \& Claussen 1990). Apparently, the dense ammonia gas in the other sources 
has been disrupted by the expansion of the \hii region and/or outflows.

\subsection{Physical conditions}

This section summarizes our knowledge of the physical conditions of the hot  
molecular gas found in the vicinity of UC \hii regions. We 
concentrate the discussion on those sources for which the physical parameters 
have been determined from observations of inversion transitions of NH$_3$ made 
with high angular resolution. The most commonly observed line is the (3,3) 
inversion transition which effectively selects the warmer and higher density 
condensations within massive cores. In Table 2 we compile most of the data 
published to date of hot ammonia cores in regions of massive star formation, 
observed with high angular resolution. The term hot ammonia core is used here 
to refer to molecular structures with either ammonia line brightness temperatures 
and/or rotational temperatures between inversion transitions of ammonia $\ge50$ K. 
This criteria ensures that the kinetic temperature of the gas is greater than 50 
K (Walmsley \& Ungerechts 1983; Danby et al. 1988). These data show that hot 
ammonia cores have ammonia column densities as high as 10$^{19}$ cm$^{-2}$, 
molecular hydrogen densities up to $7\times10^{7}$ cm$^{-3}$, and kinetic 
temperatures up to 250 K. Typically the hot ammonia emission originates from 
optically thick clumps with linear sizes of the order of 0.05 pc. 

\subsubsection{Temperatures}

The kinetic temperature of molecular gas can be estimated from the rotational 
temperature between two transitions of a molecule, which can in turn be derived 
from the ratio of the optical depths of the transitions. The ammonia molecule is 
particularly convenient for this purpose because the optical depths of their 
inversion transitions can be directly derived from the observed intensity ratio 
of their hyperfine components. Observations of two inversion transition lines of 
ammonia permits one to determine the rotational temperature. For this reason 
the NH$_3$ molecule is known as a cosmic thermometer. Strictly, however, the 
NH$_3$ rotational temperature sets a lower limit to the kinetic temperature 
(Walmsley \& Ungerechts 1983). 

The kinetic temperatures of hot ammonia cores are typically $\sim$100 K, and 
can be as high as 250 K. These temperatures are considerably higher than the kinetic 
temperatures of the $\sim$1 pc scale ambient medium (ie. massive cores) that 
harbours compact \hii regions, of typically 30 K (CWC 90). The intimate association 
between hot ammonia cores and compact \hii regions suggests that the heating 
of the molecular gas is most likely due to the strong radiation field from the 
luminous star (or stars) that excites the associated region of ionized gas. 
To be valid, this hypothesis requires that the total luminosity of the 
ammonia core be equal to or smaller than the total luminosity of the exciting star.
In particular, for cores which are offset from the associated UC \hii region their 
luminosity should be smaller  by a factor of $\Omega/4\pi$, where $\Omega$ is the 
solid angle subtended by the ammonia core from the ionizing star. The ammonia 
cores within the G32.80+0.19 and G61.48+0.09 (S88B) regions of star formation 
provide good examples of hot cores that are intimately associated with compact 
\hii regions and have luminosities that can be fully explained as a result of 
heating by the nearby star which ionizes the \hii region (G\'omez et al. 1995). 

Alternatively, hot ammonia cores may be heated by recently formed stars embedded 
at their centers and which have not yet produced a detectable ultracompact \hii 
region. Cesaroni et al. (1994a) found that the hot ammonia cores within the 
G10.47+0.03 and G31.41+0.31 massive star forming regions are offset from the 
nearby compact \hii region and have luminosities in excess of the luminosity of 
the companion HII region, suggesting that the molecular gas in these cores is 
warmed by a luminous object that is embedded within the core, and not by the star 
responsible for the ionization of the nearby compact \hii region. The estimated 
core luminosities are however larger, by factors of $\sim4-5$, than the bolometric 
luminosity derived from the IRAS data, which is thought to be an upper limit; 
an inconsistency termed  the ``luminosity paradox''. Later, Cesaroni et al. (1998) 
found that these cores are
oblate and have strong temperature gradients, and showed that the paradox 
was produced by the assumption of spherical geometry and emission 
at a single temperature. Finally, it is also possible that a non negligible 
contribution to the heating of cores could be provided by the input of mechanical 
energy due to shocks, driven either by the expansion of the \hii regions, stellar 
winds, or outflows. Since the expansion of the region of ionized gas drives a 
shock wave into the neutral gas, which sets it into motion outwards, in this 
scenario the cores are most likely to be found along the edges of the associated 
compact HII region (cf. Keto \& Ho 1989). 

The temperature across ammonia cores has been mapped in only a few cases, most of 
which appear to be surrounding UC \hii regions  (Keto et al. 1987a; Garay \& 
Rodr\'\i guez 1990; Garay et al. 1993a). These studies show a common feature: 
the presence of temperature gradients, with the temperatures being higher toward 
the associated UC \hii region. In the G10.6-0.4 molecular core the kinetic 
temperature of the gas decreases with distance from the exciting star of the UC 
\hii region, $r$, as $r^{-1/2}$, being $\sim$140 K at a distance of 
0.05 pc and $\sim$54 K at 0.35 pc (Keto et al. 1987a). In the massive core  
G34.3+0.15 the rotational temperature varies as a function of the radial distance 
as $13~ (r/{\rm pc})^{-0.6}$ (see Figure~\ref{fig-tempvsrad}), having a value 
of $\sim$185 K at the position of the ultracompact core ($r\sim$0.015 pc; Garay 
\& Rodr\'\i guez 1990). The rotational temperature of the ammonia structure 
associated with the G10.47+0.03 cluster of ultracompact \hii regions (Wood \& 
Churchwell 1989a) changes with projected distance from the exciting stars as 
$r^{-0.4}$, increasing from $\sim25$~K in the outer parts of the halo (diameter of 
$\sim0.25$ pc) to $\sim75$~K at the center of the core ($\sim0.08$ pc in size; 
Garay et al. 1993a). The observed gradients in temperature suggest that the 
molecular gas is heated via collisional excitation with hot dust, which in turn is 
heated by the absorption of radiation emitted by the central star. Scoville \& 
Kwan (1976) worked out a simple model for this process and found that the dust 
temperature, $ T_{\small D}$, at a distance $r$ from an embedded star with 
luminosity, $L_{\star}$, is 
$$ T_{\small D} = 65 \left({{0.1 {\rm pc}}\over{r}}\right)^{2/(4+\beta)}
\left({{L_{\star}}\over{10^5 L_{\odot}}}\right)^{1/(4+\beta)}
\left({{0.1}\over{f}}\right)^{1/(4+\beta)} ~{\rm K}~~, \eqno(11) $$
where $\beta$ is the power law index of dust emissivity at far infrared wavelengths
and $f$ is its value at 50$\mu$m. The power law indices of the temperature versus 
radius relationship derived from the observations are consistent with those 
predicted by this model, and suggests that $\beta$ is in the range from 0 to 1. 
In Figure~\ref{fig-tempvslum} we plot the inferred temperature at a distance of 
0.1 pc from the central exciting star (or stars) as a function of luminosity 
for the four regions with reported temperature profiles with radius (G10.47+0.03, 
G10.6-0.4, G34.3+0.15 and Cepheus-A). It shows that the temperature at a fixed 
distance from the central star increases with luminosity of the exciting star; 
a linear least squares fit to the trend gives 
$T = 39~(L_{\star}/10^5 L_{\odot})^{0.4\pm0.2}$ K. Given the error in the exponent, 
and the fact that in most of the regions considered the heating is provided by a 
cluster of stars, rather than a single star, the observed trend is consistent 
with eqn.(11) and $\beta=0$.  The above results strongly argue that 
in these objects the embedded star or stars that ionizes the UC \hii region also 
provides the bulk of the heating of the whole ammonia structures, including the hot 
cores.

\subsubsection{Column densities and masses}

Assuming that all energy levels are populated according to LTE, the total column 
density of \nhtres can be derived once the optical depth, line width, rotational 
temperature, and excitation temperature of an inversion transition are known 
(see Ho \& Townes 1983). The ammonia column densities of the compact ammonia cores 
range from $\sim 2\times10^{15}$~cm$^{-2}$ to $\sim 1\times10^{19}$~cm$^{-2}$.  
The large range of observed column densities may be due to evolutionary effects.
As the region of ionized gas grows into the molecular environment
and the molecular core expands, the NH$_3$ column density is expected to decrease.
Hence objects with the larger column densities may correspond to the youngest 
ammonia cores. 

The total mass of a spherical cloud of molecular gas can be determined from the 
derived ammonia column density, $N($NH$_3)$, and core radius, $R$, using the 
expression,
$$ M_{\small NH_3} =  \mu_g m_{\small H_2} \left[{{\rm H_2}\over{\rm NH_3}}\right] 
   N({\rm NH_3}) \pi R^2, ~~ \eqno(12) $$
where $\mu_g$ is the mean atomic weight of the gas (=1.36), and $m_{\small H_2}$ 
is the mass of a hydrogen molecule. This approach requires an assumption about the 
[{\nhtres}/H$_2$] abundance ratio, which is the largest source 
of uncertainty of the method. The [{\nhtres}/H$_2$] ratio has been estimated 
to be between {$10^{-7}$} for small, dark clouds (Ungerechts, Walmsley, \& 
Winnewiser 1980) and {$10^{-5}$} in the dense nucleus of the Orion molecular cloud 
(Genzel et al. 1982). The high temperature of the hot cores certainly increases 
the evaporation of icy mantles, making the chemical formation process in the 
vicinity of UC \hii regions different from that in cold dark clouds. Thus, the gas 
phase is expected to be enriched with mantle constituents such as NH$_3$, CH$_3$OH, 
and H$_2$O although the precise amount of enrichment is difficult to estimate. For 
warm cores in massive molecular clouds the derived fractional ammonia abundances 
are typically $\sim${$10^{-6}$} (eg. Henkel, Wilson, \& Mauersberger 1987). 

Alternatively, the mass of the ammonia cores can be computed assuming that they 
are in virial equilibrium. For a spherical core of radius $R$, the virial mass, 
$M_{vir}$, in solar masses is 
$$ M_{vir} = 210 \beta~ (\Delta v)^2 ~R ~~, \eqno(13) $$
where $\Delta v$ is the line width in \kms, $R$ is in pc, and $\beta$ is a constant, 
of order unity, which depends on the density profile of the core ($\beta=1$ for a 
uniform density core; $\beta$= 0.90 for one whose density varies as the inverse 
first power of the radius; MacLaren, Richardson \& Wolfendale 1988). 

The molecular mass of the hot ammonia cores, computed using one of the two methods 
discussed above, range from about 1 to 300 {\mo}. Although the mass estimate can 
easily differ by an order of magnitude depending on the method used and its 
underlying assumptions, and hence should be taken with caution, the wide range of 
derived masses reflects physical differences among the cores. The less massive 
($M < 30$\mo) and low density ($n$(H$_2) < 5\times10^4$ cm$^{-3}$) cores
(see Table 2), which are intimately associated with regions of ionized gas,
appear to be different from the most massive ones. 
G\'omez et al. (1995) suggest that they are probably clumps of remnant gas 
that are being heated externally by the exciting star of the \hii region. We note 
that some of the cores listed in Table 2 have been observed in other molecular and 
continuum tracers each of which might probe distinct regions of the core. In fact 
the mass estimated from millimeter and infrared continuum emission (see Kurtz et al. 
1999b) are generally larger than those derived from the ammonia observations;
the derived ratios of dust to molecular masses range from 0.5 
to 90, with a geometric average of $\sim 5$. 

\subsubsection{Densities}

The derived density of the hot ammonia cores, whose determination suffers the same 
difficulties as the mass determination, ranges from $1\times10^4$ to $7\times10^7$ 
cm$^{-3}$.  The densest hot ammonia cores are thought to correspond to the 
cradles of massive stars, whereas the less dense hot cores are thought to 
correspond to structures representing the remnant molecular core material that has 
survived the powerful effects of the formation of a luminous star. However, 
whether the high densities are produced by the dynamical interaction between the 
newly formed star and the molecular core gas or reflect the initial conditions of 
dense fragments within the cores is still an open question. In some cases
support for the first hypothesis is provided by (3,3) ammonia observations which 
show that the dense gas emission arises also from the warmer parts of the cloud 
which are found very close to the heating source. The high densities may be either 
produced by compression from shocks associated with the expansion of the \hii 
region or from bow shocks produced by the motions of the exciting star within the 
core.

The dependence of density with radius in ammonia cores has been investigated 
in a few cases. Keto et al. (1987a) found that the molecular hydrogen density in the  
G10.6-0.4 molecular core scales with radius roughly as r$^{-2}$ and is 
4$\times10^6$ cm$^{-3}$ within a centrally condensed structure of 0.05 pc in 
radius. In the G34.3+0.15 cloud the hydrogen molecular density is $\sim7\times10^7$ 
cm$^{-3}$ in the ultracompact core (size of $\sim$0.03 pc) and decreases outwards 
as $n \propto r^{-1.7}$ (Garay \& Rodr\'\i guez 1990). In the W51:e8 ammonia core 
the column density scales with radius in a region of 0.04 to 0.3 pc as 
r$^{-1.0}$, implying a density dependence of r$^{-2.0}$ (Zhang \& Ho 1997). The 
derived power law index for the density distribution falls in the range between 3/2, 
predicted for gas collapsing to a point source that dominates the gravitational 
field (e.g. Shu 1977), and 2, expected for an isothermal 
sphere in hydrostatic equilibrium or a condensation of free falling gas at 
constant speed. Whether or not these density dependences are representative of 
other massive cores remains to be investigated.

\subsection{Line widths}

The line widths of the \nhtres emission from hot cores are 
remarkably wide, with values ranging from $\sim4$ to $\sim$10 \kms. They are 
broader than the typical ammonia line width of $\sim$3 {\kms} observed
by CWC90 toward massive cores associated with ultracompact \hii regions.
We recall that the CWC90 observations sample regions of typically of $\sim$
1 pc in size which may contain hot cores.  What is the origin of the broad line 
widths in hot ammonia cores? Thermal broadening is too small to explain the 
observed line widths. Even at the high kinetic temperature of hot cores, 
$\sim200$~K, the thermal width is only $\sim$0.7 \kms. Saturation broadening, due 
to large optical depths in the lines (see Ho 1972), is also insufficient. Although 
some saturation broadening is possible in the main lines, 
which generally have large opacities, a negligible contribution is expected for 
the satellite lines which also show very broad line widths. For example, 
the observed line widths of the ammonia emission from the G10.47+0.03 hot core are 
$\sim12$ \kms in the main lines and $\sim9$ \kms in the satellite lines (Garay et 
al. 1993a). The difference in line width, of $\sim$3\kms, can be ascribed 
to saturation effects in the main lines, but the large broadening observed in the 
satellite lines requires an additional explanation. We conclude that the large 
line width of the hot cores can not be solely explained by either its high kinetic 
temperature nor by its large optical depth. Most likely, the broad line widths are 
due to the presence of ordered or turbulent gas motions, 
which are believed to consist mainly of magnetohydrodynamic waves.

Several studies have shown that a significant correlation exist between the 
observed line width and size of molecular clouds (cf. Myers 1983, and references
therein). To investigate whether or not the dense, hot, and compact molecular 
structures fulfills such a relationship we plot in Figure~\ref{fig-nh3widvsrad} 
the observed line width versus size for hot ammonia cores that are closely 
associated with compact \hii regions. For comparison also plotted are the values 
observed for dense cores in low-mass clouds (Myers 1983) and dense cores in 
massive clouds (Caselli \& Myers 1995), which are denser and warmer than the cores 
within dark clouds. Figure~\ref{fig-nh3widvsrad} clearly shows that the line width 
versus size relationship is not unique. At a given size, the hot ammonia cores 
have broader line widths than massive cloud cores which in turn are broader 
than those in low-mass clouds. The shift in the line width-size relationship 
observed for the three types of cores most likely reflects differences in the 
physical properties underlying this relation. 

The physical basis for the $\Delta v$ versus $r$ relationship is still poorly 
understood, although the prevailing view is that the line broadening is magnetic 
in origin (Shu et al. 1987; Myers \& Goodman 1988). Fuller \& Myers (1992) 
concluded that the line width versus size relation observed for low-mass cores 
within dark clouds reflects the initial conditions for the formation of low-mass 
stars rather than being a consequence of star-core interaction. Similarly, Caselli 
\& Myers (1995) concluded that the physical conditions observed in massive dense 
cores can also be considered part of the initial condition of the star formation 
process. It is then tempting to explain the observed differences in the $\Delta v$ 
versus $r$ relationship in terms of differences in the initial physical conditions 
of the different cores. 

Assuming that the three types of cores are in virial equilibrium, the differences 
in line width at a fixed cloud size could be explained as due to differences in the 
molecular densities. Approximating the observed line width-size correlations by 
power law relations $\Delta v = C r^{p}$, where $C$ and $p$ are constants 
-- different for each class of cores --, and using ecn.(13) for the virial mass, it 
follows that $n($H$_2) \propto C^2/r^{2-2p}$. For cores with sizes in the 
0.02 - 0.2 pc range (where a comparison of the ammonia data is appropriate) the 
average line width of low-mass cores, massive cores, and hot ammonia cores are 
roughly in the ratio 1:3:22, which would imply, if virialized, that their  
molecular densities are in the ratio 1:9:500.  The average densities derived from 
the observations ($\sim3\times10^4$ cm$^{-3}$ for low-mass cores (Myers 1983); 
$\sim2\times10^5$ cm$^{-3}$ for massive cores (Caselli \& Myers 1995); 
$\sim2\times10^7$ cm$^{-3}$ for the hot ammonia cores) are in the ratio 1:7:700,
close to those predicted from the observed shift in the line width-size 
relationship by the virial hypothesis. Note however that this 
simply recasts the problem of the 
shift in the $\Delta v$ versus $r$ relationship to the problem of what produces 
the high densities in hot ammonia cores. Larson (1981) suggested that the 
systematic differences in line width among the different types of cores may 
represent different amounts of evolution away from a ``primordial'' state of 
turbulence represented most closely by the low-mass cores. The more massive cores
may have experienced more gravitational contraction than the low-mass cores, which 
would increase their virial velocities. However, this hypothesis predicts that 
the line width should increase as the radius decreases as $r^{-1/2}$ which is not 
observed.  

To investigate in more detail the differences in the observed line width versus
size relationship in terms of differences in physical parameters it is convenient 
to separate the thermal and non-thermal contributions to the line width 
(cf. Myers 1983; Fuller \& Myers 1987). Caselli \& Myers (1995) found that for 
massive cores the non-thermal component of the line width, $\Delta v_{n-t}$, depends 
on size as $r^q$ with $q=0.21\pm0.03$. On the other hand, Myers 
(1983) found that for low-mass cores $q\sim0.53\pm0.07$. Caselli \& Myers (1995) 
modeled these relations in terms of cloud structure, and found that the difference 
in slopes is mainly due to differences in density structure. Cores in massive 
clouds are denser and have steeper density profiles than cores in low-mass clouds. 
We find that the non-thermal width of hot ammonia cores follows the trend 
$\Delta v_{n-t} \sim r^{0.1\pm0.1}$. Even though the amount of data is 
small, implying that errors are large, it appears significant that the slope of 
the nonthermal line width - size relationship of hot ammonia cores is shallower 
than that of cores in massive clouds. It is thus tempting to suggest that the 
shallower slope of the hot ammonia cores imply that they are denser and 
have steeper density profiles than massive cores. 

In the above discussion it was assumed that hot ammonia cores are in 
hydrostatic equilibrium, being supported against collapse by a non-thermal
mechanism most likely of magnetic origin (Myers \& Goodman 1988). In this picture, 
the hot cores might correspond to the basic units of massive star 
formation. The assumption of hydrostatic equilibrium is debatable, however. As 
discussed in the preceding sections, the hot ammonia cores are likely to be 
battered by the winds and heated by the radiation from the luminous star that 
excites the nearby \hii region. The broad line widths of the hot cores may then 
reflect molecular gas that is undergoing outflow motions, possibly driven by the 
expansion of the hot ionized gas around the newly formed stars. The hot ammonia 
gas would then be part of the heated and compressed region behind the shock front 
around a compact \hii region. Clearly in this case the line widths can not be used 
to derive virial masses. Whether the physical properties of the hot ammonia cores 
correspond to those of the initial process of the formation of massive stars or 
have been strongly affected by the presence of the intense UV fields and winds 
from the nearby stars remains to be investigated. On the other hand the hot 
ammonia cores could be undergoing collapse. Interestingly, models of collapsing 
hot molecular cores assuming that they collapse from an initial singular
logatropic density distribution can account for the large observed non-thermal 
line widths (Osorio, Lizano, \& D'Alessio 1999; see \S 5.2.2). 

\subsection{Kinematics}

In this section we discuss the results of observations, made with high angular 
resolution, of the velocity field toward hot cores which permit one to 
resolve the structure of the motions across them, and hence allow one to 
investigate the dynamics of the hot circumstellar molecular gas. A range of gas 
motions, involving either rotational, infalling or outflowing 
motions, have been observed. 

\subsubsection{Infall}

The first firm detection of an accretion flow of molecular gas onto newly 
formed massive stars was found in the case of G10.6-0.4, the brightest member 
of a complex of compact \hii regions. From observations with moderate angular 
resolution ($9''$), Ho \& Haschick (1986) found a rapidly rotating core 
embedded within a more extended, lower density, slowly rotating envelope. 
In addition, part of the molecular gas within the core is accreting onto the 
central \hii 
region. The later result was confirmed by high angular resolution observations 
(0\pas3) which show that the molecular gas close to the central \hii region 
exhibits differential rotation and accelerating infall (Keto, Ho, \& Haschick 
1987a, 1988). These studies strongly suggest that the molecular gas within the 
core of G10.6-0.4 is gravitationally collapsing and spinning up as it approaches 
the central star. Similar spectroscopic signatures, indicating collapse and 
rotation motions possibly associated with the original accretion flows, were 
observed by Keto, Ho, \& Reid (1987b) toward the W3(OH) and G34.3+0.15 regions of 
active star formation. All these cores have embedded UC \hii regions
showing that the central star has arrived on the main sequence stage
while the natal molecular gas is still undergoing gravitational collapse.

W51 is other high-mass star forming region where infall motions have been 
extensively studied. Rudolph et al. (1990) proposed that the entire W51 
molecular cloud complex is in a state of gravitational collapse involving 
$4\times10^4$\mo. However, interferometric observations show that the contraction 
is confined to small ammonia condensations of only a few tenths of a parsec 
(Ho \& Young 1996). The ammonia cores located near the ``W51e'' cluster of UC 
\hii regions, which are excited by B0 - B0.5 stars (Gaume, Johnston, \& Wilson 
1993), exhibit strong indications of infall and rotational motions (Zhang \& Ho 
1997; Zhang, Ho, \& Ohashi 1998a). As in the case of the sources mentioned above, 
the W51:e2 core is in a stage where the central star has already arrived on the 
main sequence with the surrounding gas still contracting onto the mass 
concentration at the center (Ho \& Young 1996). 

\subsubsection{Expansion}

Ammonia emission from dense molecular condensations associated with young massive 
stars indicating expansion motions have been reported in several cases (eg. Vogel
et al. 1987, Keto \& Ho 1989, G\'omez et al. 1991, Wood 1993, Garay et 
al. 1993a). In most of them the expansion motions are ascribed to the emergence of 
an expanding \hii~ region which sweeps up the remnant ammonia gas into thin 
filaments and clumps. For example, toward the W33-Main region the NH$_3$ emission 
appears immediately surrounding regions of ionized gas and their velocity structure 
indicates that the shells of molecular gas are expanding away from the \hii regions 
(Keto \& Ho 1989). These results suggest that the molecular emission arises from 
the compressed and heated gas behind the shock front driven by the expansion of 
the \hii regions.

Alternatively, the expansion motions of warm ammonia structures may be produced 
by the presence of powerful outflows in the region. Toward the G5.89-0.39 
UC \hii region, which is associated with a rapid (total velocity of $\sim$60 \kms)
bipolar outflow traced by OH maser features (Ziljstra et al.~1990) 
and with a more extended, massive, and energetic flow observed in CO (Harvey \& 
Forveille 1988), the ammonia observations show the presence of a massive 
($\sim$30\mo) and warm ($\sim90$K) molecular envelope that is undergoing 
expansion (G\'omez et al. 1991, Wood 1993). Similarly, toward the DR21 star 
forming region the ammonia line emission traces hot shocked gas, corresponding 
to the remnant material that has survived a powerful outflow (Wilson et al. 1995). 

\subsubsection{Rotation} 

Ammonia structures within massive star forming regions exhibiting the presence 
of rotational motions have been reported in a few cases. Most of these structures 
have flattened morphologies and diameters greater than 0.1 pc, corresponding to 
interstellar, rather than circumstellar, disks. Jackson, Ho, \& Haschick (1988) 
detected a molecular disk or toroid, of $\sim$0.3 pc in diameter, associated with 
the compact \hii region NGC 6334 F (Rodr\'\i guez, Cant\'o, \& 
Moran 1982). The disk, of $\sim30$\mo, is gravitationally bound and rotating 
about a central star of $\sim$30\mo~ that excites the compact \hii region.  
Gaume \& Claussen (1990) found that the ammonia condensation within the Sgr B2 Main 
\hii complex (Vogel et al. 1987) is intimately associated with an ultracompact 
region of ionized gas (source F) which shows a shell or ring-like morphology. The 
ammonia core has an extent of $\sim$0.1 pc and exhibits a velocity gradient, 
roughly along a north-south direction, of $\sim$25 \kms across the source. They 
suggest that the ammonia structure delineates a molecular disk or torus, 
surrounding the Sgr B2 F \hii~ region, that is rotating with a velocity of $\sim$11 
\kms, implying a mass interior to the disk of $\sim$1400\mo.

The first firm evidence from ammonia observations of the presence of a rotating 
circumstellar disk was presented by Zhang, Hunter, \& Sridharan (1998b) who found 
that the flattened disk-like structure around the young high-mass star IRAS 
20126+4104 (${\cal L} \sim 1.3\times10^4$ \lo), exhibits a velocity gradient across 
the major axis. The motions of the disk, which has a radius of $\sim5\times10^3$ AU,
are consistent with Keplerian rotation around a 20 \mo~ point mass. 

\subsection{Evolutionary states} 

Vogel et al. (1987) proposed that hot cores are produced during a 
relatively brief stage after the star begins to heat the surrounding medium. At a 
very early stage, dense material is still falling toward the star while in the 
later stages, marked by the emergence of an expanding \hii region, the dense gas 
is dispersed by outflows. In addition, as current theories of star formation 
assert (cf. Shu et al. 1987), one also expects to see rotating 
circumstellar disks. Thus, by studying the core kinematics, it should be possible 
to identify ammonia cores over a range of evolutionary stages, from cores 
undergoing accretion to condensations experiencing outflows. The observations 
described in the previous section permit one to identify an evolutionary sequence 
within observed ammonia cores. In the remaining of this section
we incorporate in the discussion results obtained not only 
from ammonia observations but from other tracers of hot cores.

\subsubsection{Collapse phase} 

The existence of cores in a collapsing phase has been well established 
observationally (see \S 3.5.1). The presence of infalling gas toward a central mass 
concentration is demonstrated by position velocity diagrams which show the 
classic ``C'' or ``O'' shapes consistent with radial motions projected along the 
line of sight. The available infrared and radio data even allow one to distinguish 
different stages within the collapse phase. Ammonia cores with infall motions 
and with embedded UC \hii regions, implying that a massive star has already 
formed at their centers, most likely mark the oldest stage of the collapse 
phase. Explanations in terms of pure free fall collapse of the molecular gas 
toward the central star (or stars) is difficult, since the free fall collapse 
time of the core, {$\sim 10^{4}$} years, is short compared with the timescale of 
formation of the luminous early type OB stars at the center of the flow 
($\sim10^{5}$ years). It has been suggested (e.g. Ho \& Haschick 1986) that the 
molecular gas in the core is spiralling toward the compact \hii regions, slowly 
falling but rapidly spinning-up due to partial conservation of the angular 
momentum. 

Cores in an earlier evolutionary stage are distinguished by being luminous
but not associated with an UC \hii region (eg. Cesaroni et al. 1994a; Hunter et 
al. 1998; Molinari et al. 1998). They are characterized by having densities of 
$\sim 10^7$ cm$^{-3}$, sizes $L \sim 0.1$ pc, and temperatures $T > 100$ K. 
Walmsley (1995) proposed that these cores host inside a young massive OB-type 
star (or stars) that is still undergoing an intense accretion phase. 
The high-mass accretion rate of the infalling material quenchs the development 
of an UC \hii region (Yorke 1984), and the free-free emission
from the ionized material is undetectable at centimeter wavelengths. The mass 
accretion rate is also large enough that the ram pressure of the infall could 
provide the force to prevent the expansion of an \hii region. The minimum accretion 
rate needed to choke off an incipient \hii region, $ \dot{M}_{crit}$, is 
$$ \dot{M}_{crit} = 2 \sqrt{\pi} \mu m_{\small H} \left({{N_u G M_*}
\over{\alpha_B}}\right)^{1/2}~~,  \eqno(14) $$
where $M_*$ is the mass of the central object,
$\mu$ is the mean atomic weight, and $G$ is the gravitational 
constant. We will refer to this type of objects as collapsing hot molecular cores 
(CHMCs) to distinguish them from other types of hot molecular cores. Osorio et al. 
(1999) modeled the spectrum of several CHMCs as massive envelopes accreting onto a 
young massive central B type star (see \S 5.2.2). In order to fit the data, 
accretion rates in the envelopes $\dot M >5\times10^{-4} M_{\odot}$ 
yr$^{-1}$ are required, implying that the main heating agent is the accretion 
luminosity. The lifetime of these objects is short $\lesssim 10^{5}$ yr. 
The central star is expected to increase its mass by accretion up to $M_* \sim 30-40 
M_\odot$ until the radiation pressure onto dust grains reverses the infall in 
timescales of $\sim 10^4$ yr.  Osorio et al. (1999) suggest that radiation 
pressure together with a stellar wind can help clear out the accreting flow. 
An UC \hii region is then produced; its evolution being governed by the reversal 
of the accretion flow.

The fraction of the observed ammonia cores that are in the CHMC stage is debatable. 
Most of the luminous ammonia cores without radio emission reported so far have been 
found in the neighborhood of \hii regions, and it is not clear how much of its heating 
is due to internal or external sources of energy. Recently, Kaufman, Hollenbach, 
\& Tielens (1998) calculated the temperature structure of dense hot cores with either 
internal or external illumination. They find considerable differences in the 
temperature structure; internally heated cores can more easily produce large column 
densities ($N_{H_2} \ge 10^{23}$ cm$^{-2}$) of hot ($T > 100$ K) gas than externally 
heated cores.  These results are important because they can be used to 
observationally discern between different types of hot cores. 

A potential candidate of a molecular core in this evolutionary stage is object 
W51:e8 in W51, which show spectroscopic characteristics of infall of a centrally 
condensed cloud (Ho \& Young 1996, Zhang \& Ho 1997). Radio continuum emission is 
undetected at 3.6 cm at a sensitivity of $<$ 1 mJy (Gaume et al. 1993) and detected 
($\sim$ 17 mJy) at 1.3 cm wavelengths. The spectral index is much stepper than the 
one expected for an optically thick \hii region suggesting that the flux density 
at 1.3 cm is dominated by dust emission (Zhang \& Ho 1997). The lack of free-free 
emission associated with this object could be interpreted as due to the action of 
accreting matter which keeps the region of ionized gas close to the stellar radius 
and prevents its expansion. Strong evidence for the presence of a luminous 
molecular core that may be the precursor of an UC \hii region is presented by 
Molinari et al. (1998), who found that the luminous (${\cal L} \sim 1.6 \times10^4 
$\lo) IRAS source 23385+6053 is coincident with a 3.4 mm compact and massive (M 
$\sim$ 470 \mo) core, but is undetected in the radio continuum and at mid-IR 
($\lambda < 15 \mu m$) wavelengths. This core would then correspond to a 
``forerunner'' of an UC \hii region.

Finally, ammonia cores with inward motions implying the presence of a 
massive object, but underluminous at infrared wavelengths and without detectable 
free-free emission, would represent cores in the earliest possible collapse stage, 
contracting towards a central mass condensation but without yet giving rise to 
a luminous protostar. 
Compact ammonia cores that exhibit radial flows and that are not associated with 
detectable sources of radio continuum emission have been observed within the star 
forming regions DR 21 (Keto 1990), G10.5+0.0 (Garay et al. 1993a), and NGC6334 
(Kraemer et al. 1997). Their kinetic temperatures are consistently smaller than 
those of ammonia cores associated with compact \hii regions. To show that these 
objects are effectively in the earliest evolutionary stage of infall, undergoing 
gravitational collapse and accretion just prior to the formation of a massive 
protostar, and not in the CHMC phase, it is crucial to determine their infrared 
luminosity with high angular resolution. 

The positive identification of a protostellar condensation after gravitational 
collapse has formed a massive stellar core but before nuclear reactions 
have set in remains one of the main goals of observers of young massive stars. 
Even though direct evidence of the process of collapse of a dense and compact 
molecular core leading to the formation of a single massive star has yet to be 
obtained, there is an increasing number of objects that may be tracing very 
early stages of massive star formation, as evidenced by their association to 
H$_2$O masers, bipolar molecular outflows, and/or warm dust emission 
but no detectable, or weak, thermal emission from ionized gas (eg. objects 
G75.78+0.34:mm and NGC6334F:mm; Carral et al. 1997). Interestingly, several of 
these objects have been found located near the head of cometary UC \hii regions 
(Hofner \& Churchwell 1996; Carral et al. 1997). Other luminous ($L_{bol} > 10^3$ 
\lo) objects identified with massive protostellar objects without detectable levels 
of radio continuum free-free emission are G34.24+0.13MM (Hunter at al. 1998) and 
the TW object (Wilner, Welch, \& Foster 1995) which is also associated with \hdo 
masers, a bipolar outflow and a jet of synchrotron emission (Reid et al. 1995).

\subsubsection{Expansion phase} 

The evidence for the existence of a late stage in which the ammonia condensations 
are undergoing outflowing motions is also compelling and has been presented in 
\S 3.5.2. The outflowing motions are produced either by the expansion of the 
recently formed \hii region or by the presence of molecular outflows.
In several cases the hot ammonia emission appear surrounding the newly formed
UC \hii region suggesting compression and heating by the ionization-shock 
front at the boundary of the \hii region. The shell-like structure of molecular 
clumps around the \hii regions may represent the latest stage in the evolution 
of a warm molecular core, before it is disrupted by the expansion of the \hii 
region. This hypothesis is supported by the fact that the densities of the ammonia 
cores projected toward \hii regions are generally lower than those found 
located beyond UC \hii regions. Thus, OB stars play an important role in the 
heating, increase in velocity dispersion, and finally in the destruction of
the massive cores that gave them birth.

\subsubsection{Initial phase}

The first phase of the evolutionary sequence in the formation of a massive star,
representing the initial condition for the actual phase of gravitational collapse,  
is thought to consist of a dense and compact molecular core. We will refer to 
this structure as a high-mass pre-stellar core. Unlike its counterpart in sites 
of low-mass star formation, namely the ammonia cores (Myers \& Benson 1983), 
its physical characteristics are still unknown although it must contain a mass 
of at least 20\mo~ within a region with a size $\le$ 0.05 pc (McCaughrean \& 
Stauffer 1994; Hillenbrand 1997). An important issue which has not yet been 
discerned is the temperature of such a massive pre-stellar core, namely the 
temperature just before massive star formation.

Although different strategies have been applied in search of these objects, the 
inherent quiescent massive pre-stellar cores have been difficult to detect. A 
number of searches have been made using as targets UC \hii regions. Given the 
tendency of O and B stars to form in clusters and the continuing process of star 
formation within an association, it is expected to find molecular structures that 
will give rise to the next generation of massive stars near UC \hii regions. 
In fact, most of the known hot ammonia cores have been identified using this 
criteria and thus a strong selection effect is associated with the current 
available sample. The nearby OB stars have enough radiative and mechanical 
luminosity to substantially alter the initial core conditions. Such an environment 
is thus considerably impacted by the presence of the existing massive stars, and 
whether or not the physical conditions of the hot cores near UC \hii regions 
represent the initial conditions of massive pre-stellar cores is debatable (cf. 
Stahler, Palla, \& Ho 1999). Hot cores either contain a central star or have been 
heated by nearby massive stars and therefore their physical characteristics are 
unlikely to reflect initial conditions. It is possible that at the expected high 
densities and low temperatures 
of high-mass pre-stellar cores molecules are frozen on dust grains and are 
significantly depleted in the gas phase, so that their detection can not be easily 
achieved via molecular line observations. Thus, others have sought for compact, 
massive condensations of cold dust emission based on high spatial resolution 
observations in the millimeter and submillimeter wavelength range. In particular, 
Mezger et al. (1988) found half a dozen compact dust emission knots with masses 
between 10 and 60\mo~ embedded in the NGC 2024 star forming region. 
Mezger et al. (1988) derived that the knots have high densities 
($n_H\sim10^8-10^9$ cm$^{-3}$), sizes from 0.003 to 0.03 pc, and dust temperatures 
of $\sim16$K, and suggested that they correspond to isothermal massive protostellar
condensations without luminous stellar cores. However, observations 
of CS($7\rightarrow6$) line emission showed that the dust knots are closely 
associated with molecular condensations with gas temperatures of $\sim$ 45 K, 
inconsistent with the 16 K derived from the dust emission, suggesting that the 
submillimeter continuum peaks contain heavily obscured luminous stellar cores 
(Moore et al. 1989). Another strategy adopted in the search for the molecular 
precursors of high-mass stars is to use as targets \hdo maser sources that are not 
associated with radio continuum sources (cf. Codella \& Felli 1995). As will 
be discussed in \S 4.1 water masers are powerful signposts of regions of massive 
star formation and their detection almost guarantees the presence of high-mass 
stars. The most promising approach in the search of massive pre-stellar cores, 
and determination of their physical parameters and dynamics, will be 
through high angular resolution observations at submillimeter wavelengths in 
tracers of cold, high density molecular gas, which is one of the main goals of 
the next-generation submillimeter arrays.  

\vfill\eject

\section{MASER EMISSION}

Since the discovery of maser phenomenon in the interstellar medium 35 years ago 
(Weaver et al. 1965), several studies have shown that three molecules exhibit 
widespread and powerful maser emission in regions of active star formation: 
H$_2$O, OH, and CH$_3$OH. More recently, maser emission in molecular lines of 
H$_2$CO and NH$_3$ has been detected toward a few compact \hii regions (Hofner 
et al. 1994; Pratap, Menten, \& Snyder 1994; Kraemer \& Jackson 1995). Besides 
acting as powerful signposts of active star formation, the intense 
maser emission provides a unique tool to probe the physical conditions and 
kinematics of these regions on scales of $10 - 10^3$ AU. The purpose of this 
section is to shortly summarize what has been learned about the dynamics of these 
regions and the spatial and physical association to other tracers of massive star 
formation. For excellent reviews of maser theory and observations see Reid \& 
Moran (1981, 1988) and Elitzur (1982).
   
\subsection{H$_2$O}

Interstellar water masers are characterized by appearing in small clusters
of features or centers of activity, with extents of $\sim$10-100 AU, and by 
exhibiting a wide spread in velocity, of $\sim 50-100$ \kms (Genzel \& Downes
1977; Genzel et al. 1978). From a survey of water emission in the 6$_{16}-5_{23}$ 
rotational transition at 22 GHz toward compact \hii regions, 
Churchwell et al. (1990) found that the frequency of ocurrence of water masers in 
the neighborhood of newly formed O and B stars to be $\sim$67\%, demonstrating that 
water maser action is a common phenomenon in regions of star formation. \hdo masers 
have been found in young forming regions containing stars as massive as O7 down to 
at least B3, their luminosity being proportional to the total FIR luminosity of 
the region (Moran 1990; Palagi et al. 1993). However, the detailed relationship 
between the \hdo masers and individual stars in these regions has yet to be 
established. Hofner \& Churchwell (1996) found that water masers associated
with \hii regions of cometary morphology always appear in clumps located near, yet 
offset from, the cometary arcs. In most cases the \hdo centers of activity are 
either undetectable at radio continuum wavelengths or associated to weak radio 
continuum sources. 
In some cases the water masers are associated with hot molecular cores,
suggesting that they might mark the position of young protostellar objects.

Using VLBI techniques it is possible to study the \hdo centers of activity 
with milli arcsec resolution allowing measurements of proper motions and radial 
velocities, thus providing velocity information in three dimensions. However, 
very few observations of this kind have been made so far. Most of the \hdo masers 
studied with VLBI reveal radial expansion from a single origin (Orion-KL: Genzel et 
al. 1981; Sgr B2: Reid et al. 1988), implying that the H$_2$O maser phenomenon is 
associated with mass outflows. Using the VLA, with angular resolution of 0.08", 
Torrelles and collaborators have recently mapped the water 
maser emission toward three regions of star formation containing multiple O and 
B stars, radio continuum jets and molecular outflows. They found that the \hdo 
masers exhibit a dichotomy in their spatial distribution with respect to that 
of the radio jet: they either trace outflows (masers oriented along the major axis 
of the radio jet) or disks around YSOs (masers distributed perpendicular to the 
major axis of the radio jet). The water masers in disk-like distributions appear 
to surround a B star, and their motions indicate either rotation (W75[B], NGC2071; 
Torrelles et al. 1997, 1998a) or a combination of rotation and contraction 
(Cepheus A; Torrelles et al. 1996, 1998b). These results are very important since 
they show the presence and yield the kinematics of disks at scales as small as 
10 AU. The masers found along the continuum jet exhibit a close spatial and 
velocity correspondence with the outflowing molecular gas, showing that they are 
taking part in the bipolar outflow.

\subsection{OH}

Hydroxyl masers are known to be excellent indicators of UC \hii regions.
Surveys of OH maser emission, in the 1.665 GHz line, toward star forming complexes 
show that the OH masers are always found associated with the 
densest and most compact \hii region within the complex. The OH maser emission 
arises from spots with linear sizes of $\sim10^{14}$ cm and are distributed 
typically over regions of $\sim10^{16}$ cm. The OH maser spots are generally 
found projected on the face of the compact \hii region (e.g. Garay, Reid, \& 
Moran 1985; Gaume \& Mutel 1987). Garay et al. (1985) found that 
in most cases the OH masers are redshifted from the velocity of the 
associated UC \hii region, and suggested that the OH maser 
probably lie in a remnant accreting envelope, outside the advancing ionization 
front of the \hii region, that is still collapsing toward the recently formed 
star. However, in the case of W3(OH) the proper motions of the OH maser spots
suggest that they are in an expanding molecular shell between the shock front 
and the ionization front from the \hii region (Bloemhof, Reid, \& Moran 1992).  

\subsection{CH$_3$OH}

Fifteen years after the discovery of methanol maser emission (Barrett, 
Schwartz, \& Waters 1971) it was recognized that maser action in methanol lines 
is a common phenomenon in regions of massive star formation (Menten et al. 1986). 
Numerous masing transitions of methanol have been observed at cm and mm 
wavelengths.  The strongest masers at cm wavelengths arise from the J$_2$-J$_1$ 
series of E type methanol at 25 GHz (Barrett, Ho, \& Martin 1975; Menten et al. 
1986); the $2_0-3_{-1}$E transition at 12.2 GHz (Batrla et al. 1987, Norris et 
al. 1987); the $2_1-3_0$E and $9_2-10_1$A$^+$ transitions at 19.9 and 23.1 GHz 
(Wilson et al. 1984, 1985); and the $5_1-6_0$A$^+$ transition at 6.7 GHz (Menten 
1991; McLeod \& Gaylard 1992; Caswell et al. 1995a). In the mm regime strong 
emission has been found in the $5_{-1}-4_0$E transition at 84 GHz (Batrla \& 
Menten 1988); the $8_0-7_1$A$^+$ transition at 95 GHz (Plambeck \& Menten 1990);  
the $3_1-4_0$E transition at 107 GHz (Val'tts et al. 1995), and the J$_0$-J$_{-1}$ 
series of E at 157 GHz (Slysh, Kalenskii, \& Val'tts 1995).
   
Two types of methanol masers have been identified (Batrla et al. 1987, Menten 
1991): Class I masers which show maser emission in the $4_{-1}-3_0$E,
$7_0-6_1$A$^+$, $5_{-1}-3_0$E, $8_0-7_1$A$^+$  and J$_2$-J$_1$ E transitions, 
and Class II masers which show maser emission in the $2_0-3_{-1}$E, $2_1-3_0$E, 
$5_1-6_0$A$^+$, $9_2-10_1$A$^+$, $7_{-1}-8_{-1}$E, $6_2-5_3$A$^{-}$ and 
$6_2-5_3$A$^{+}$ transitions . None of the lines that mase in Class I sources 
also mase in Class II sources and vice versa.  Class I sources are found in the 
general vicinity of massive star forming regions, but are offset from compact 
\hii regions, strong infrared sources, OH maser centers, and other indicators of 
the formation of high-mass stars. Plambeck \& Menten (1990) suggested that Class 
I masers are associated with shock fronts, indicating the interface of
interaction between mass 
outflows and dense ambient material. Alternatively, Class I masers may be 
associated with earlier phases of stellar evolution and may indicate the true 
protostars which are still accreting mass. In contrast, Class II masers are closely 
associated with OH interstellar masers and compact \hii regions (Menten et al. 
1988; Norris et al. 1988; Caswell, Vaile, \& Forster 1995b). They are frequently 
found spatially distributed along lines or arcs, suggesting that they might be 
associated with either shock fronts, jets, or edge-on protoplanetary disks around 
the nascent star (Norris et al. 1988; 1993).

\subsection{NH$_3$ }

The first detection of maser emission in the ammonia molecule was 
presented by Madden et al. (1986), who reported maser emission in the 
(J,K)=(9,6) inversion transition toward four regions of active star formation
(W51, NGC7538, W49, and DR21[OH]). Since then, NH$_3$ maser emission 
from the W51 region has been reported in several other inversion transitions 
(Mauersberger, Henkel, \& Wilson 1987; Wilson \& Henkel 1988; Wilson, Gaume, \& 
Johnston 1991; Zhang \& Ho 1995, 1997), and detected toward the G9.62+0.12 and 
NGC6334 star forming regions (Hofner et al. 1994, Kraemer \& Jackson 1995).

The relationship between the NH$_3$ maser action and the birth of young massive 
stars has yet to be established. Pratap et al. (1991) found that the 
ammonia masers towards the W51e \hii region are closely associated to, although not 
coincident with, the two ultracompact \hii regions W51e1 and W51e2.
On the other hand, toward the G9.62+0.12 \hii complex the NH$_3$ maser is 
coincident with a weak continuum source and a water maser (Hofner et al. 1994). 
Although the number of sources studied is small, the NH$_3$ masers exhibit the 
same dichotomy in their spatial distribution as the H$_2$O masers, tracing either 
disks or outflows.  For instance, in the NGC7538 region the NH$_3$ masers are 
coincident with the core of the bipolar compact \hii region, show a velocity 
gradient, and seem to arise in a thick rotating molecular disk surrounding IRS1 
(Gaume et al. 1991). Toward NGC6334 I, the NH$_3$ masers are located at the end 
of high velocity bipolar outflows suggesting that they are produced by 
shocks and mark the location where the molecular outflow jet impinges upon the 
ambient medium.

\subsection{Relationship between masers}

The discussion presented in the previous sections show that masers are frequently 
associated with bipolar outflows, infrared sources, and UC \hii regions, 
demonstrating that they are closely associated with the early stages of O and B 
star formation. The physical association between masers and other signposts of 
high-mass star formation has however yet to be established. In general there is 
a good overall positional coincidence between the masers and dense molecular clumps 
seen in thermal emission lines (cf. Gaume et al. 1993). On the other hand, in the 
complex regions of star formation where maser emission has been detected in 
different molecular species, the maser spots do not appear to be coincident with 
each other. A good illustration of this is presented in Figure~\ref{fig-maser} 
which plots the position of \hdo, OH, CH$_3$OH, and NH$_3$ maser spots toward the 
G9.62+0.19 massive star forming region (Hofner et al. 1994), showing that the maser 
spots are usually not coincident. These authors propose that the striking alignment 
of masers, hot molecular clumps, and UC \hii regions, along a linear structure of 
$\sim$0.6 pc, is produced by a shock front, associated with the expansion of nearby 
more evolved \hii regions, that has compressed the molecular gas and triggered star 
formation along the front. 

The lack of positional coincidence between different maser species suggests that 
they might indicate different evolutionary stages in the formation of massive stars 
(e.g. Lo, Burke, \& Haschick 1975; Genzel \& Downes 1977). The fact that \hdo masers 
are found located near, but not coincident, with compact radio continuum sources, 
whereas OH masers often coincide with UC \hii regions, might indicate that \hdo 
masers are generally associated with stellar objects at an earlier stage of 
evolution than OH masers. It is thought that \hdo masers appear in the earliest 
stages of protostellar evolution, being excited in a high density medium close to 
a deeply embedded protostar. They may arise from either a disk around a 
central protostar or from condensations or density enhancements in the outflow 
from a young stellar object which has not yet ionized a region large enough to 
be detectable. Methanol masers, which also seem to be signposts of disks around 
young massive objects, are in most cases associated with a detectable radio 
continuum source. This suggests that methanol masers indicate a later evolutionary 
stage than \hdo masers, in which the central protostar is currently developing a 
detectable UC \hii region. OH masers, which are always associated with UC \hii 
regions, appear last. They exist in a compressed, possibly infalling, shell of 
molecular gas just outside the Str\"omgren sphere of a very young star. The whole 
maser phenomenon ocurrs during the earlier phases of evolution of massive stars and 
fades away as the \hii region evolves into the diffuse stage (Codella et al. 1994b).

The relationship between masers and the sequence of maser events asserted above 
may not be unique, however. In some cases there is a close association between 
several types of masers and UC \hii regions, which has been attributed to the 
presence of shocks. An example of this is the positional agreement between the 
\hdo, OH, and NH$_3$ masers and the ultracompact radio source E in G9.62+0.19, 
which is thought to correspond to a partially ionized stellar wind (Hofner et 
al. 1996). The masers toward this deeply embedded object are thought to mark 
an extremely young massive star in the process of destroying the remnant 
molecular material from which it formed. The CH$_3$OH masers near this source are 
displaced from the center of activity defined by the other masers, and are probably 
excited by the collisions of dense clumps in the wind with clumps in the ambient gas. 

\vfill\eject

\section{THE FORMATION OF MASSIVE STARS}

According to current ideas the first stage in the formation of massive OB stars
is initiated by contraction and fragmentation of a giant molecular cloud 
leading to the formation of dense molecular cores in approximate hydrostatic 
equilibrium. This premise is supported by observations which show that GMCs are 
composed of numerous dense cores and have small volume filling factors (eg. Blitz 
1993). The cores appear to have a considerable range of physical parameters. The 
distribution of core mass follows a power law relation with an index of $-1.6$ 
(see Blitz 1991; and references therein), close to the value of $-1.5$ predicted 
as the result of an equilibrium between coagulation and fragmentation (Spitzer 
1982; Genzel 1991). Little is known, however, about the relationship between 
the mass spectrum of the cores and the initial mass function of the massive stars. 
The most massive stars appear to form in the dense cores of massive clouds
and the mass of the most massive young star increases systematically with 
the mass of the associated molecular cloud (Larson 1982). In addition, the average 
star formation efficiency increases with cloud mass (Williams \& McKee 1997). 

The discussion presented in the preceding sections raises a host of questions 
related to the structure of molecular cores and the formation of massive stars, 
such as: Once a molecular core forms, how does it evolve to produce massive stars?
What are the roles of fragmentation and accretion? Do the more massive cores 
lead to a simultaneous formation of a group of massive stars or does the formation 
of massive stars proceed individually from well separated dense smaller clumps? 
Is there a basic unit within dense cores that will eventually produce a single 
massive star? How often does the collapse of massive molecular cores lead to the 
formation of clusters of massive stars rather than to a single high-mass star?
Are accretion disks and bipolar outflows present in the formation of massive 
stars? Although no definite answer to these questions have been given yet, 
partial answers are beginning to emerge. In the following discussion we first 
summarize the main observational properties that have been gathered about the 
formation of massive stars, then discuss basic theoretical principles of a few 
physical phenomena that are relevant for the development of a theory of massive 
star formation, and finally speculate on a possible scenario for the formation 
of massive stars.

\subsection{Observational characteristics}

\subsubsection{Clustering, gregariousness, and segregation}

A wealth of radio continuum observations of deeply embedded star forming regions 
in our Galaxy have shown that massive stars are gregarious, namely that they tend 
to be born in groups (e.g. Ho \& Haschick 1981; Welch et al. 1987; Rudolph et al. 
1990; Garay et al. 1993b). As discussed in \S 2.4, Garay et al. (1993b) found that 
most of the luminous IRAS point sources associated with compact regions of ionized 
gas exhibit complex structure in their radio continuum emission, suggesting that the 
complex morphologies are produced by excitation from multiple stars and thus 
implying that a cluster of OB stars have been formed. The radio free-free 
observations can only detect massive, luminous stars within the region and thus they 
do not provide information regarding the less luminous non-ionizing stars.

The ability to investigate the population of newly formed stars deeply embedded 
in molecular clouds has recently improved with the advent of infrared array 
detectors. Several regions containing high 
mass stars have been extensively studied: W3 (Megeath et al. 1996), Mon R2 
(Carpenter et al. 1997), Orion-Trapezium (Hillenbrand \& Hartmann 1998), NGC 2024 
(Lada et al. 1991), giving valuable clues to the mechanism of 
formation of massive stars. These observations show that massive stars do not 
form in isolation but are instead born in relatively rich clusters. The young 
embedded clusters have typically stellar volume densities of $\sim10^4$ 
stars/pc$^3$ and sizes of 0.2-0.4 pc. Furthermore, they provide clear evidence 
that the most massive stars within the dense cluster are preferentially formed 
near its central region. The concentration of OB stars near the center of young 
stellar clusters is unlikely to be the result of dynamical evolution since the 
ages of the stars are smaller than the required collisional relaxation times 
(Hillenbrand \& Hartmann 1998; Bonnell \& Davies 1998). In addition, the 
gravitational potential of the cluster still has a significant contribution from 
molecular gas that is usually thought to impede the process of 
dynamical mass segregation. The preference for high-mass stars to be located in 
dense, central cluster regions is thus likely to be a primordial feature, namely 
an indication of where massive stars are formed. The mass segregation in young 
stellar clusters would then reflect their formation scenario and the cluster 
initial conditions.

\subsubsection{Bipolar molecular outflows}

Bipolar molecular outflows represent an important evolutionary stage in the 
formation of low-mass stars (see Shu 1991 and Bachiller 1996 for reviews on 
theory and observations, respectively). Although systematic studies to identify 
bipolar outflows from massive young stellar objects (YSOs) have begun only 
recently, there is growing evidence that bipolar molecular outflows are also 
associated with massive stars (Snell et al. 1984; Scoville et al. 1986; Garden et 
al. 1991; Harvey \& Forveille 1988; Bachiller \& Cernicharo 1990; Kastner et al. 
1994; Shepherd \& Churchwell 1996b; Acord, Walmsley, \& Churchwell 1997; Shepherd, 
Churchwell, \& Wilner 1997), showing that this phenomenon is also a basic component 
of the formation process of massive stars. To investigate the frequency of 
occurrence, Shepherd \& Churchwell (1996a) searched for high-velocity (HV) line 
wings toward $\sim 120$ high-mass star forming regions and found that the presence 
of HV gas is quite common ($\sim90 \%$). If the HV gas is due to bipolar outflows, 
then these results indicate that molecular outflows are a common property of newly 
formed massive stars.

A list with most of the currently known massive bipolar outflows detected toward 
young massive star forming regions and their properties has been recently compiled 
by Churchwell (1999). The bipolar outflow masses range from about 10 to 4800 \mo, 
with a mean value of 130 \mo; the mass outflow rates range from $3\times10^{-5}$ 
to $3\times10^{-2}$ \mo~ yr$^{-1}$, and the kinetic energies range from 
$1\times10^{46}$ to $6\times10^{48}$ ergs. Whereas the dynamical ages of outflows 
in massive star forming regions are similar to those associated with low-mass YSOs, 
their masses, outflow rates and energetics are substantially larger (Churchwell 
1997). In particular, the mean mass and luminosity of outflows in massive star 
forming regions are roughly a factor of 100 times greater than 
those from low-mass stars (Churchwell 1999). These results indicate that the 
luminous outflows are driven by massive stars which are indeed more energetic and 
hence are able to inject more energy into their surroundings than low-mass stars. 
In fact, it has been found that the mass outflow rate, force, and mechanical 
luminosity of the molecular outflows are tightly correlated with the stellar 
luminosity of the driving source (Cabrit \& Bertout 1992; Shepherd \& Churchwell 
1996b).

Although most of the massive outflows are found in the immediate vicinity of UC 
\hii regions, the identification of the driving source is not straightforward 
since in many cases, if not always, a cluster of emerging massive stars is 
present in these regions. Possibly the best way to identify the energy source 
would be through observations of the radio continuum emission from the stellar 
jet that is expected to be associated with the driving source (e.g. Rodr\'\i guez 
et al. 1994). We note that a few massive outflows do not show the presence of 
compact sources of radio continuum emission that could be identified as the 
driver of the bipolar outflow. In these cases the driver of the flow is probably 
a massive protostar that has not yet been able to form a detectable \hii region 
because of rapid mass accretion (see \S 3.6.1). Particularly notable are the 
luminous objects IRAS 20126+4104 ($L_{bol} \sim 1.3 \times 10^4$ \lo) and IRAS 
23385+6053 ($L_{bol} \sim 1.6\times 10^4$ \lo) which are associated with 
outflows, are luminous in the millimeter wavelength range, but are undetected 
in the continuum at centimeter wavelenghts (Cesaroni et al. 1997; Molinari et al. 
1998), suggesting that they are bona fide massive class 0 objects, namely massive 
protostars in a very early accretion phase (Andr\'e, Ward-Thompson, \& Barsony 1993). 

The observational results described above are of invaluable importance, 
implying that bipolar molecular outflows are a basic component of the formation 
process of all stars. However, the large masses and energetics associated 
with the luminous outflows raise several questions which have not yet been answered, 
such as: What is the origin of the mass in massive outflows? and What is the 
driving mechanism of the luminous outflows? 
Although the physics of the high-mass outflows remain to be addressed, it 
has been suggested that a common driving mechanism could operate across the 
entire mass or luminosity range (Richer et al. 1999).

\subsubsection{Disks}

Most of the flattened structures detected within high-mass star forming regions 
have diameters in the range 0.1 -- 1 pc, corresponding to interstellar, rather 
than circumstellar, disks (Vogel \& Welch 1983; Jackson et al. 1988, Gaume \& 
Claussen 1990; Nakamura et al. 1991; Ho, Terebey, \& Turner 1994; Kraemer et al. 
1997). A large fraction of these elongated structures exhibit smooth velocity 
gradients, suggesting that they correspond to rotating disks. Their masses range 
from 10 to 2000\mo. In some cases rotation in combination with collapse or 
expansion motions are detected. The massive rotating interstellar disks may provide 
the natural environment for the formation of a cluster of stars. Whether they 
correspond to dynamically bound disks is not yet known.

The evidence for the presence of circumstellar disks ($< 10^4$ AU) around 
young massive stars was, until recently, null. However, sensitive, high 
angular resolution observations of dust and molecular gas emission are beginning 
to change this state of affairs. Synthesis observations of the SiO maser 
emission from IRc2, a young star with a luminosity of $\sim 10^{5}$ \lo) in the 
Orion K-L region, provide strong evidence for the presence of an 80 AU diameter 
rotating and expanding circumstellar disk (Plambeck, Wright, \& Carlstrom 1990). 
Dust emission from this disk has been detected at middle IR wavelenghts (7.8 and 
12.5$\mu$: Lester et al. 1985; 3.6$\mu$: Dougados et al. 1993). An elongated dust 
feature of $\sim 10^2$ AU is detected at near infrared wavelenghts toward 
the luminous ($\sim3\times10^4$ \lo) star MWC349A (Lienert 1986). Zhang et al. 
(1998b) found that the flattened disk-like structure around the young 
high-mass star IRAS 20126+4104 (${\cal L} \sim 1.3\times10^4$ \lo), first 
detected by Cesaroni et al. (1997) in CH$_3$CN, exhibits a velocity gradient 
across the major axis. The kinematics of the disk, which has a radius of 
$\sim5\times10^3$ AU, are consistent with Keplerian rotation around a 
20 \mo~ star.

A new and exciting probe that is providing distinct evidence of the presence of
circumstellar disks rotating around a massive OB star is 
methanol maser emission. Interferometric observations towards a few southern 
massive star forming regions show that the CH$_3$OH masers lie along lines or 
curves and have linear velocity gradients (Norris et al. 1998). The line of 
masers within source G339.88-1.2, a compact \hii region excited by a B0.5 star, 
lies across the diameter of the \hii region suggesting that they are tracing a 
circumstellar photoevaporated disk (Ellingsen, Norris, \& McCulloch 1996). This 
has been confirmed by observations at 10 $\mu$ which show a disk-like dust 
feature aligned with the methanol masers
(Stecklum et al. 1998). In sources G309.92+0.48 and G336.42-0.26 the motions 
of the methanol maser spots are essentially Keplerian implying the presence of 
bound disks, with diameters of $4\times10^3$ and $3\times10^3$ AU, around 
central masses of 10 and 14 \mo, respectively (Norris et al. 1998). 
Interestingly, Natta, Grinin, \& Mannings (1999) found that Herbig Be stars,
which are in a later stage of evolution than the above objects, are not 
associated with the compact continuum emission in the millimeter expected from 
circumstellar disks, 
suggesting that the disks may have been eroded by the strong radiation fields.

\subsection{Physical processes}

The sequence of processes leading to the formation of massive stars from the 
parental molecular cloud is not yet well understood, and is one of the most 
important challenges in the field of star formation. Clustered star formation is 
a complex hydrodynamic problem which in addition to gravity, angular momentum, 
magnetic fields, and heating and cooling, should include feedback effects from 
dense proto-stellar cores and newly formed stars.

Since the formation of massive stars occurs simultaneously on a number of 
gravitating centers, implying some kind of global synchronization of the process, 
a comprehensive theory of massive star formation must necessarily provide a 
thorough account of the progression of molecular gas over several decades in 
spatial scales (see Larson 1999). It should yield a detailed description of all 
stages of evolution, beginning with the initial stage in which the large scale 
fragmentation of a molecular cloud, containing clumps and dense filaments, leads 
to the formation of dense massive cores, proceeding through intermediate stages in 
which the fragmentation of dense cores leads to the formation of multiple 
pre-stellar cores, and ending with the final stage of evolution of a protostar 
which accretes from its surrounding dense core, and possibly interact with other 
pre-stellar cores through winds, radiation and/or dynamically, leading to the 
formation of a massive star. The stages of building stars from the collapse of 
their parent molecular clouds is thought to involve two main processes: 
fragmentation and accretion. In what follows we discuss the importance 
of these processes in forming massive stars.

\subsubsection{Fragmentation}

Fragmentation processes are believed to be important in the initial stage of 
the collapse of large clouds of molecular gas. The details
of the mechanism of gravitational fragmentation are poorly known.
Theoretical studies indicate that gravitational fragmentation during
three-dimensional collapse is a highly inefficient process, but that fragmentation 
will occur in flattened or filamentary structures (Layzer 1963; Larson 1972, 1985). 
Fragmentation can also become significant if large density perturbations are 
present initially, the small perturbations being damped out during the collapse 
(eg., Tohline 1980a,b). 

Numerical simulations of the fragmentation of an isothermal collapsing cloud
show that condensations can indeed be formed on a dynamical time scale as a
result of gravitational fragmentation (e.g., Larson 1978; Monaghan \& Lattanzio
1991). In particular, Larson (1978) found that the number of condensations
formed in the collapse of an isolated cloud is comparable to the number of
Jeans masses initially present. For a cloud with gas temperature, $T$, and 
molecular hydrogen density, $n_{H_2}$, the Jeans mass, $M_J$, representing the 
minimum mass of gas that is required for gravitational collapse is   
$$ M_J =  5.6 \left({{T}\over{10~K}}\right)^{3/2} 
 \left({{n_{H_2}}\over{10^4 cm^{-3}}}\right)^{-1/2}~ M_{\odot}~, $$
which decreases with increasing density. To assess 
the validity of Larson's result under different initial conditions, Klessen, 
Burkert, \& Bate (1998) followed the isothermal collapse and fragmentation of a 
small highly gravitationally unstable region, consisting of a hierarchy of clumps 
and filaments. They find that the large scale density fluctuations within the region 
under consideration collapse on themselves and into filaments and knots in about a 
free fall time of the isolated region. Dense cores are formed in the center of the 
most massive Jeans-unstable clumps, soon followed by the collapse of clumps of 
lower initial mass, altogether creating a hierarchical structured cluster of highly 
condensed cores. The mass distribution of the resulting cores peaks at about twice 
the initial Jeans mass in the system, and ranges from 0.3 to 15 Jeans masses. Thus, 
at the end of the fragmentation phase a fragment's mass is typically of the order 
of the initial Jeans mass, roughly independent of the initial conditions.

Nevertheless, a wealth of observations show that giant molecular clouds and
massive cores are supported against gravity by turbulent motions and magnetic 
fields rather than by thermal pressure (McKee 1989; Bertoldi \& McKee 1992).
Thus, the concept of Jeans mass may be of little use in the definition of
the appropriate initial conditions in the context of fragmentation of
molecular clouds. Furthermore the above results, whereby a collapsing cloud 
fragments into clumps with masses of the order 
of the initial Jeans mass, do not easily explain the formation of massive stars. 
Massive stars appear to form in the dense ($n_{H_2} \sim 3\times 10^5$ cm$^{-3}$) 
cores of massive cold clouds where the Jeans mass is only $\sim 1.0$ \mo, 
considerable smaller than those of OB stars. This conclusion rests, 
however, on the assumption that the cloud temperature is similar to that observed 
in dark clouds producing low-mass stars, $T\sim10$ K. It has been argued that more 
massive fragments could form if the gas temperature in high-mass star forming 
regions is higher. As discussed in \S 3.3.1, temperatures of 100-200 K are now 
routinely measured toward hot cores near \hii regions suggesting that fragments with 
masses of the order of a massive star can be produced in such environments. 
However, hot cores might be or are influenced by the presence of a nearby, or 
embedded, massive star, whose large luminosity raises the temperature of the 
ambient dust and gas. Thus, it is not clear if this temperature is that  
before the formation of massive stars. It might be possible that the 
temperature of massive cores increases gradually with time as energy is injected 
in the surrounding medium by low and intermediate mass stars formed during an 
initial stage of star formation, and that the massive stars are formed last, 
during a late ``hot'' stage of star formation. Nevertheless, we stress that 
the effects of magnetic fields and turbulence at large scales in molecular
clouds will dominate the cloud structure. In particular, their support allows
the clouds to achieve much higher densities than with thermal pressure 
support alone before becoming gravitationally unstable (Caselli \& Myers 1995;
Crutcher 1999).

Another mechanism of fragmentation might occur in turbulent flows. Based on 2D 
hydrodynamical simulations of the ISM at kpc scales, Ballesteros-Paredes, 
V\'azquez-Semadeni \& Scalo (1999) proposed that GMCs and their substructures 
form as density fluctuations induced by large scale interstellar 
turbulence due to colliding streams of gas (see also V\'azquez-Semadeni, Passot, 
\& Pouquet 1996). However, in these simulations quasi-hydrostatic configurations 
like the observed ammonia cores have not yet been produced.

\subsubsection{Coalescence and accretion}

It is believed that accumulation processes, such as interactions among clumps 
or forming stars (e.g. Larson 1978, 1982, 1990), and/or accretion processes play a 
crucial role in the formation of massive stars. If most of the gas in a 
massive molecular core is in clumps, a clump may gain mass through coagulation with 
other clumps or through accretion from residual gas. Larson (1992) suggested that 
core coalescence can account for the observed power-law of the stellar initial mass 
function (see also Silk \& Takahashi 1979). A key issue is to identify a mechanism 
that increases the efficiency of accretion processes in dense proto-cluster cores. 
Larson (1982) suggested that tidal forces from an association of stars produce
the disruption of incipient condensations within the remaining gas with the gas 
settling toward the center of the forming cluster and becoming progressively more 
dense. Much of this dense gas may then give rise to a few massive stars. Bonnell 
et al. (1997) investigated the effects of accretion of gas in a cloud containing 
initially a number of nucleating centers, and found the accretion process to be 
highly non-uniform, with a few nucleating centers accreting significantly more than 
the rest. A large dynamic range in the final core masses is obtained, with the 
nucleating structures near the cloud center accreting, due to their location at 
the bottom of the potential well, most of the material and hence giving rise to 
the most massive stars.  
There are then two possible stages of accumulation of gas: an initial 
stage of coalescence in which dense cores grow by accretion and interactions with 
other cores and become progressively more massive and condensed, and a 
final accretion stage in which a protostar mainly accretes from its surroundings. 

Recently, Stahler et al. (1999) suggested that OB stars 
could form inside dense clusters by coalescence of already existing 
stars of lower mass. Since the cross section for these type of encounters
is too small for naked stars, they suggested that the coagulation occurs 
when the low-mass stars are still surrounded by dense molecular cores
which increases the effective cross sections. 

An alternative scenario for the formation of massive stars is direct accretion 
onto a central object from a surrounding pre-stellar dense, massive ($M \ge 100$ 
\mo) core. This accretion scenario assumes the existence of such structures but do 
not address the problem of their formation. A potential problem in forming massive 
stars by accretion from an infalling envelope is the difficulty of accreting once 
the protostar has attained a mass $\ge 10$ \mo. The large radiation pressure on 
dust grains, produced by the intense radiation field of such a protostar, can halt 
the collapse and reverse the infall (Larson \& Starrfield 1971; 
Kahn 1974; Yorke \& Kr\"ugel 1977). This problem may be circumvented in two ways. 
First, the above limit applies to accretion in a spherically symmetric collapse. 
However, the pre-stellar core is likely to have some angular momentum resulting 
in the formation of a protostellar disk. Accretion onto the massive protostar may 
then proceed from the disk. In this case, Jijina \& Adams (1996) found that the 
maximun luminosity to mass ratio for radiation pressure on dust grains to reverse 
the infall is much less restrictive. Second, Wolfire \& Cassinelli (1987) 
showed that for inflows with mass accretion rates $\dot{M} \ge 
10^{-3}$\mo~ yr$^{-1}$  the ram pressure is sufficiently strong to overcome 
the radiative forces on dust by the luminous central star, allowing continuous 
accretion to form massive stars.

Recently, Osorio et al. (1999) developed models of collapsing hot 
molecular cores assuming that the cores collapse from an initial singular 
logatropic density distribution characterized by a logatropic equation of state 
$P = P_0 \ln (\rho/\rho_0)$, where $\rho_0$ is an arbitrary reference density and 
$P_0$ the pressure at this density. This type of structure has been invoked by 
several authors (Lizano \& Shu 1989; Myers \& Fuller 1992; McLaughlin \& Pudritz 
1996) to explain the linewidth-density relation observed in molecular clouds 
($\Delta v \propto \rho^{-1/2}$; e.g. Larson 1981) since in logatropes the sound 
speed, $c_s$, given by $c_s = [d P / d \rho]^{1/2} = [P_0 / \rho]^{1/2}$, depends 
on density like the observed relationship. The mass weighted quadratic velocity 
dispersion of a logatropic core is 
$$c_s^2 = \left\{{{8 \pi G P_0}\over{9}}\right\}^{1/2}  R_{\rm core} F(x)~~,$$ 
where $ R_{\rm core}$ is the core size and $F(x) = (1+x+x^2)/(1+x)$, where 
$x = s/R_{\rm core}$ is the fraction of the core size that is actually probed by 
observations in a given molecular line. The distance $s$ is measured from the
outer boundary of the core, $R_{core}$, towards the center of the core. For an 
optically thin transition, $x \sim 1$ and $F(x) \simeq 3/2$, whereas for an 
optically thick transition, like $NH_3$ lines, $x << 1$ and $F(x) \simeq 1$. 
This velocity dispersion corresponds to the virial speed that provides the 
hidrostatic support of the logatropic core. Appreciable free-fall speeds are 
expected only at scales of few tens of AU from the core center.
The expected line width (FWHM) averaged over the source is  
$$\Delta v = 6.8 \left({P_0 \over 4\times10^{-7} ~ {\rm dynes~ cm}^{-2}}\right)^{1/4}
\left( {R_{\rm core} \over 0.1 {\rm pc}}\right)^{1/2} [F(x)]^{1/2}~~ \kms,$$
where the normalization factors in $P_0$ and $R_{\rm core}$ are typical 
of the cores modeled by Osorio et al. (1999). Thus, this type of models 
can account for the large line widths observed in hot molecular cores as shown 
in Figure~\ref{fig-nh3widvsrad}. 

\subsubsection{Magnetic fields}

Magnetic fields are believed to play an important role in the formation process 
of stars. They may initially provide the support of molecular clouds against 
gravity (e.g. see reviews of Heiles et al. 1993; McKee et al. 1993).
A molecular cloud can be supported against its self-gravity by magnetic fields 
alone provided its mass is less than the magnetic critical mass, $ M_{\Phi}$, 
$$ M_{\Phi} =  {1 \over 2 \pi} {\Phi \over G^{1/2}}~~, $$
where $\Phi (= \int B dA)$ is the magnetic flux threaded by the cloud (Mestel 
\& Spitzer 1956; Mouschovias \& Spitzer 1976; Tomisaka, Ikeuchi \& Nakamura 1988; 
1989; Li \& Shu 1996). 
For a cloud with radius $R$ and magnetic field of strength $B$, $\Phi \simeq 
\pi~ R^2 B$. If the cloud mass is greater than the critical mass, $M > M_{\Phi}$, 
the cloud is called supercritical. It has been suggested that, in the absence 
of other substantial means of support, supercritical clouds will collapse to 
form a closely packed group of stars. Instead, subcritical clouds with  $M < M_{\Phi}$,
will condense locally as the magnetic field support is lost via ambipolar 
diffusion. This dichotomy is the basis of the proposed mechanism of bimodal
star formation in which ``loosely aggregated'' versus ``closely packed''
star formation occurs (Shu et al. 1987; Shu et al. 1993).
  
Since molecular clouds are lightly ionized, the predominantly neutral (molecular) 
matter feels the Lorentz force through collisions with ions that slip past the 
neutrals in the process known as ambipolar diffusion, first studied by Mestel \& 
Spitzer (1956). This causes a local weakening of the magnetic support that allows 
the condensation of dense cloud cores with subcritical masses. The evolution of 
axisymmetric clouds and 
dense cores to a centrally condensed state via ambipolar difusion of the 
magnetic field has been investigated by several authors (e.g. Nakano 1979, 1982; 
Lizano \& Shu 1989; Tomisaka, Ikeuchi \& Nakamura 1990; Basu and Mouschovias 
1994; Ciolek \& Mouschovias 1994). As a result of the loss of magnetic field 
support, 
the cloud core reaches a ``gravo-magneto catastrophe'' where the central density
tries to reach infinite values (Shu 1995). To describe the transition between 
quasistatic evolution by ambipolar diffusion and dynamical gravitational collapse,
Li \& Shu (1996) introduced the idea of a pivotal state with a scale free, 
magnetostatic, density distribution approaching $\rho \propto r^{-2}$, for an 
isothermal equation of state, when the mass-to-flux ratio has a constant value, 
a condition they termed ``isopedic''. These pivotal states flatten 
as the degree of magnetic support is increased. Although small dense cores 
that give rise
to low-mass star formation are effectively isothermal, the situation may be 
different for larger and denser regions that yield high-mass or clustered star 
formation. Recently, Galli et al. (1999) have generalized these pivotal states
to self-gravitating isopedic clouds with a polytropic equation of state 
with negative index $n$. These polytropic equations of state try to mimic 
the support due to the observed nonthermal motions in molecular clouds. In 
particular, the associated sound speed increases with decreasing density as 
the observed linewidth-density relation for molecular clouds (see \S 3.4). 
These pivotal states may represent the initial states for dynamical collapse
to form a star or group of stars.

\subsection{Collapse models}

Theoretical studies of gravitational collapse, intended to model the collapse 
of a dense molecular cloud to form a star, have been made by several authors. 
In particular, the problem of finding self-similar solutions for the gravitational 
collapse of isothermal spheres has been investigated by Larson (1969), Penston 
(1969), Shu (1977), Hunter (1977), Whitworth \& Summers (1985), and  Foster \& 
Chevalier (1993). Larson (1969) and Penston (1969) studied the collapse of an 
isothermal sphere with initial uniform density which evolves to a central region 
of homologous inflow. In their similarity solution, the central flow tends to a 
free-fall collapse with velocity proportional to $r^{-1/2}$ and density proportional 
to $r^{-3/2}$, whereas in the outer parts the inflow velocity is constant at 3.3 
times the 
sound speed and the density is proportional to $r^{-2}$. On the other hand, Shu 
(1977) derived a self-similar solution for the collapse of a singular isothermal 
sphere with an initial density distribution of the form
$$ \rho =  {{A a^2}\over{4\pi G}} {{1}\over{r^2}}~~,  $$
where $a$ is the local sound speed, and A is a dimensionless constant. If A=2 
the sphere is initially in hydrostatic equilibrium and the cloud collapse begins 
at the center and spreads outward. In this case the mass accretion rate 
is constant in time and equal to $\dot{M} = m_0 {{a^3} \over{G}}~, $
where $m_o$ is a constant (=0.975). In the Larson-Penston (LP) solution, the 
initial accretion 
rate onto the central object is $\sim$47 times higher than in Shu's solution, and 
declines monotonically. Hunter (1977) and Whitworth 
\& Summers (1985) showed that there is a continuum set of similarity solutions 
for the collapse of an isothermal sphere; the LP (A=8.85, m$_0$=46.9)
and Shu (A=2, m$_0$=0.975) solutions being the extreme cases (the fastest and 
slowest collapse, respectively). 
In particular, models of the condensation of cloud cores via ambipolar diffusion
discussed in the previous section do show the tendency of the core's density to
acquire a scale free profile $\rho \propto r^{-2}$.

Models that include the effects of magnetic fields have been investigated by 
several authors. Galli \& Shu (1993a,b) followed the collapse of a magnetized 
isothermal cloud and found that strong magnetic pinching forces deviate the 
infalling gas to the equatorial plane to form a flattened disequilibrium structure 
around the star (a ``pseudo-disk''). Tomisaka (1995; 1996) and Nakamura, Hanawa 
\& Nakano (1995) studied the collapse and fragmentation of magnetized cylindrical 
clouds and found that geometrically thin disks perpendicular to the symmetry 
axis are formed which evolve toward a central singularity.  Shu \& Li (1997) 
found that in isopedic thin disks, the magnetic tension exerts a force that acts 
to dilute the disk self-gravity, while magnetic pressure is proportional to the 
gas pressure that gives support to the disk in the vertical direction. They found that 
fragmentation of the disk does not occur, but instead the disk collapses 
inside-out to form a single compact object (Li \& Shu 1997; see also Nakamura 
et al. 1995).  Chiueh \& Chou (1994) considered the collapse of a spherical 
magnetized cloud taking into account only the isotropic pressure of a tangled 
magnetic field, and found that the collapse proceeds in an inside-out fashion, 
as in the unmagnetized case. A similar approach was followed by Safier, McKee \& 
Stahler (1997) who investigated the quasi static evolution of spherical magnetized 
clouds due to ambipolar diffusion of the magnetic field and the subsequent inside-out 
collapse of the envelope to a central mass, ignoring magnetic tension and thermal 
pressure. Li (1998a) extended the work of Safier et al. (1997) to include thermal
pressure and followed the evolution of the core through the formation
of a point mass at the center (see also Li 1998b). Ciolek \& K\"onigl (1998) 
calculated the evolution through the central point mass formation including 
the magnetic tension (see 
also Contopoulos, Ciolek \& K\"onigl 1998). Recently, Allen \& Shu (1999) followed 
the self similar collapse of the isothermal pivotal states studied by Li \& Shu 
(1996).  All these calculations show that for magnetized clouds, 
the mass accretion rates increase with respect to that in the unmagnetized case. 
This happens because of the higher densities supported by magnetic fields in 
the equilibrium state, and also, as found by Galli \& Shu (1993b), because the 
slowing down of the infall speed by magnetic forces is largely offset by the 
increase of the signal speed at which the wave of infall propagates outward.

\subsection{Star formation}

Shu's solution provided the basis for the development of the current paradigm 
of the gravitational collapse of low-mass cores leading to the formation of 
low-mass stars (Shu et al. 1987; Shu et al. 1993). In this 
model a self-gravitating core initially supported by 
magnetic fields gradually contracts and becomes centrally condensed as the 
magnetic field is lost via ambipolar diffusion producing an unstable 
isothermal core with an $r^{-2}$ density distribution. Finally gravity 
predominates over magnetic forces and an inside-out dynamical collapse occurs. 
The collapse begins at the center of the core and propagates outwards at the 
effective sound speed, $a_{eff}$,
$$ a_{eff} = (1+2\alpha+\beta)^{1/2} \left({{k~T}\over{\mu m_{H_2}}}\right)^{1/2}~~, 
 \eqno(15) $$
where $\alpha$ and $\beta$ are the ratios of the magnetic pressure ($B^2/8\pi$),
and turbulent pressure $(\rho<\delta v^2>)$ to thermal pressure ($\rho kT/m$),
respectively, producing a density distribution of $r^{-3/2}$ close to the central 
point mass, behind the expansion wave. The rate, $\dot{M}$, at which the central 
object accumulates matter is,
$$ \dot{M} = 0.975 {{a_{eff}^3} \over{G}}~. \eqno(16) $$
As a result of the initial rotation of the core a flattened disk is produced at 
the center 
of the collapsing structure (c.f. Terebey, Shu, \& Cassen 1984). There follows 
an outflow stage in which the protostar deposits linear momentum, angular momentum, 
and mechanical energy into its surroundings through jets and molecular outflows,
while still accreting mass. Finally, the protostar settles onto the ZAMS. The 
predictions of this model are in good agreement with observations of collapse in 
several low-mass cores (e.g. Zhou et al. 1990) and has become the paradigm of low 
mass star formation. Nonetheless, Mardones (1998) and Tafalla et al. (1998) recently 
claimed that the strong molecular line asymmetries found in dense cores around low 
mass YSOs require infall speeds at large distances from the 
YSOs, which, even though subsonic, are inconsistent with pure inside-out collapse 
models. On the other hand, the theory of gravitational collapse of high-mass cores 
leading to the formation of massive stars has not yet been well developed. The recent 
observational results are providing valuable constraints for theoretical models.

\subsection{Formation Scenario}

The initial evolution of molecular clouds to form dense pre-stellar massive cores 
is perhaps the most poorly understood aspect of massive star formation (cf. Larson 
1999). Also, the dynamical processes taking place within massive cores to form 
massive stars are yet to be understood. The observational evidence gathered during 
the last decade suggests that in clouds with masses above a certain value, the 
formation of a cluster is favored instead of isolated star formation. In particular, 
the largest ($\sim$ few pc) massive cores seems to give rise to stellar clusters 
and OB associations. The simultaneous process of formation of massive stars on a 
number of gravitating centers is thought to be related to the initial stage of 
gravitational fragmentation which leads to the synchronized formation of pre-stellar 
cores. Magnetic fields are thought to play an important role in the support of the 
cloud before collapse, allowing high densities in the cores to be achieved.

Some authors have argued that the first stage in the formation of massive stars 
is the overall collapse of a massive core. 
A spectacular case where large-scale, seemingly organized, massive star formation 
has taken place is that of the central region of the W49 A massive molecular cloud 
core (Welch et al. 1987). W49A exhibits a remarkable ring of compact \hii regions, 
lying within a region of 2 pc in diameter, made of ten distinct ionized regions 
each containing at least an O star. The ring is apparently rotating with an 
angular velocity of 13 \kms~pc$^{-1}$, implying a total mass within the ring of 
$\sim5\times10^4$ \mo. In addition, it appears that over a spatial extent of 
$\sim$3 pc the molecular gas is moving toward the \hii regions, both on the near 
and far side of the ring. Welch et al. (1987) suggest that the rotating 
ring of UC \hii regions indicates a cluster of OB stars that was formed by 
gravitational fragmentation of the flattened rotating structure associated with 
the collapse of a single molecular cloud. Other star forming region in which 
the presence of overall gravitational collapse on large size scales involving 
large ($> 10^4$ \mo) masses have been proposed are G34.3+0.15 (Carral \& Welch 
1992; Heaton et al. 1993), W51 (Rudolph et al. 1990), and G5.89 (Wilner 1993). 
The hypothesis of global collapse in W49A has, however, been challenged by 
Serabyn, G\"usten, \& Schultz (1993) who instead suggest that the enhanced O star 
formation has been triggered by a collision of two clouds composed of several 
clumps (see also Mufson \& Liszt 1977). In addition, De Pree et al. (1997) found 
that the UC \hii regions in the ring do not have systematic velocities as a 
function of position, the individual regions appearing to be associated with 
distinct molecular clouds observed toward this region. 

It is probable that the early evolution of a massive core involves the gravitational
contraction of subregions, accompanied by the dissipation of turbulent motions, 
to finally produce the observed dense hot cores. Nakano's (1998) suggestion that
the decay of turbulence in molecular clouds may trigger the formation of stars is 
supported by recent numerical simulations of MHD waves which show that 
magnetohydrodynamic turbulence is short lived and must be constantly replenished 
(MacLow et al. 1998; Stone, Ostriker, \& Gammie 1998). Generalized subsonic motions 
observed toward low-mass cores have been explained as due to decay of turbulence 
(Myers and Lazarian 1998). Recently, Shu et al. (1999) proposed that either 
ambipolar diffusion or turbulent decay will produce scale free centrally condensed 
pivotal states. These pivotal states may correspond to the massive pre-stellar cores 
and will dynamically collapse to form massive stars. 

The new observational evidence presented in the previous sections
show that accretion disks and molecular outflows appear to be intrinsic 
to the formation process of high-mass stars. A list of massive young objects 
associated with both disks and outflows is given in Table 3. 
This evidence suggests that 
massive stars are formed in a similar manner as low-mass stars, although in a 
medium of much larger density. We propose that the formation process of massive 
stars from pre-stellar cores, left by a still not well understood combination of   
accretion, turbulent and fragmentation processes, is analogous to that of 
low-mass stars. The clumps giving rise to massive stars are, however, considerably 
denser and possibly hotter than the dark cores giving rise to low-mass stars. 
Besides the shorter time scale of the dynamical process involved in the collapse 
of a massive star, a key difference is in the mass accretion rate. Mass accretion 
rates as high as $6\times10^{-3}$ \mo yr$^{-1}$ have been estimated in collapsing 
cores associated with high-mass star forming regions (eg., W51:e2; Zhang \& Ho 1997),
whereas those associated with the formation of low-mass stars are typically 
$\sim10^{-6}$ \mo yr$^{-1}$. The problem posed by the radiative forces on dust 
halting the accretion inflow can be overcomed through the high accretion rates 
(Wolfire \& Cassinelli 1987). Assuming that all the infalling mass reaches the 
stellar surface, the time scale to form a 30\mo~ star under this infall rate is 
about $5 \times 10^3$ yr. The large values of the mass accretion rates are 
consistent with the predictions of the inside-out collapse model for a medium with 
a large amount of initial hydrostatic support, such as in the logatropic cores 
(see \S 5.2.2), and with the large observed mass outflow rates. Using the typically 
observed line widths of the ammonia emission of $\sim7$ \kms (FWHM), implying 
initial turbulent or Alfvenic velocity dispersion of $\sim3$ \kms, eqn.[16] predicts 
that $\dot{M}$ should be of the order of $6\times10^{-3}$ \mo yr$^{-1}$.

\section{SUMMARY}

We reviewed the results of recent high spatial resolution observations of 
free-free emission and molecular line emission made toward regions of high-mass 
star formation which have significantly contributed to the understanding of the 
physical conditions, morphologies, and dynamics of the ionized and molecular gas in 
the immediate vicinity of recently formed massive stars. The main results that are 
relevant for the study of the formation process of massive stars are summarized as 
follows.

The radio continuum observations of UC \hii regions show that massive stars are 
formed in clusters. The recombination line observations, which show broad profiles 
and that the line width increases as size decreases, imply that winds are present 
from a very early stage in the evolution of newly formed stars. They also provide 
definitive evidence for the presence of collimated ionized bipolar outflows, 
and indirect evidence of the presence of circumstellar disks.  

The observations of ammonia thermal emission show the existence of dense 
($>10^5$ cm$^{-3}$), hot ($>$ 50 K), and small ($<0.1$ pc) structures embedded 
within larger, less dense, and cooler structures of molecular gas. These hot ammonia 
cores are invariably located near and in several cases intimately associated with 
UC \hii regions. The observed kinematics and derived physical conditions 
of the hot molecular cores have permitted to establish an evolutionary sequence 
among them. Hot cores in the earliest stage of evolution are undergoing 
gravitational collapse, with mass accretion rates as high as $6\times10^{-3}$ 
\mo~yr$^{-1}$. They exhibit densities up to a few $10^8$ cm$^{-3}$, 
suggesting that massive stars form in regions of molecular clouds of exceptionally 
high density, high temperatures, up to $\sim250$ K, and masses of typically a few 
$10^2$ \mo. Their luminosities imply that they are heated by a recently formed 
massive star embedded at their centers, but due to the high-mass accretion rates 
they have not yet produced a detectable UC \hii region.
Hot molecular cores in the latest stage of evolution are intimately associated 
with UC \hii regions and are undergoing outflowing motions, which might be 
produced either by the expansion of the recently formed \hii region or by the 
presence of molecular outflows. These cores, whose densities and masses are lower 
than those in early stages and whose luminosities are consistent with heating 
by the star exciting the UC \hii region, are likely to represent remnant structures 
of molecular gas before being disrupted by the expansion of the \hii region. 

The observations of methanol and water maser emission show the presence, within 
dense and hot molecular cores, of linear molecular structures with Keplerian 
rotation, which provide the most direct evidence of the existence of circumstellar 
disks around young massive OB stars. The disks may provide intense mass accretion 
and play an essential role in the formation of massive stars. 

The observations of thermal CO emission show the presence of collimated bipolar 
outflows associated with massive stars, implying that this 
phenomenon is also a basic component of the formation process of massive stars. 
The bipolar outflows driven by massive stars are considerably more massive, 
luminous, and energetic than those associated with low-mass stars.
The ionized bipolar outflows associated with UC \hii regions
are likely to mark a late stage of the massive outflow phenomenon.

To summarize, the observational evidence discussed in this review, showing the 
existence of hot and very dense molecular structures undergoing large mass 
accretion rates, and attesting that during the process of collapse of massive stars 
massive disks are formed and that bipolar outflows appear, indicate that the 
formation of massive stars from massive pre-stellar cores share similar 
characteristics to those of low-mass stars. Although the massive pre-stellar 
cores appear in clusters and could be formed by a complex accumulation
process starting with smaller clumps within massive clouds, the new 
observational evidence suggest that massive stars are formed by the collapse of 
single massive pre-stellar cores, rather than by an accumulation process, and 
suggest that the paradigm of low-mass star formation is more universal than 
previously thought.  

\acknowledgments

We are very grateful to our colleagues E. Churchwell, R. Larson, D. 
Mardones, J. Moran, and  L.F. Rodr\'\i guez for many stimulating discussions 
and comments on the manuscript.
G.G. gratefully acknowledges support from a Chilean Presidential Science 
Fellowship and from the Chilean Fondecyt Project 1980660. S.L. gratefully 
acknowledges the support of the John Simon Guggenheim Memorial Foundation, 
of CONACyT and DGAPA/UNAM.

\clearpage

\figcaption[]{ 
Relationship between physical parameters of compact and ultracompact 
\hii regions. Top: Electron density versus diameter. Bottom: Emission measure 
versus diameter. The lines are least squares linear fits to the data points
(squares: Wood \& Churchwell 1989a; circles: Garay et al. 1993; 
pentagons: Gaume et al. 1995; triangles: De Pree et al. 1997).
\label{fig-neemvsdiam}}

\figcaption[]{
Ionizing photon rate versus diameter relationship for compact and ultracompact 
\hii regions (stars: Kurtz et al. 1994; all other symbols same as in Figure 1).
Marked on the right axis is the number of ionizing photons emitted by 
ZAMS stars with spectral types from B2 to O4. 
\label{fig-nivsdiam}}

\figcaption[]{
Morphologies of ultracompact \hii regions. Top right: Radio continuum emission 
of the bipolar source NGC 7538 (Campbell 1984). Top left: Radio continuum emission 
of the  shell source G45.07+0.13 (Turner \& Matthews 1984). Bottom: Radio 
continuum emission of the cometary source G34.3+0.15 (Gaume et al. 1994). 
\label{fig-hiimorph}}

\figcaption[]{
Line width versus diameter relationship for \hii regions. Triangles: extended 
\hii regions (Garay \& Rodr\'\i guez 1983); squares: typical \hii regions 
(Churchwell et al. 1978); stars: compact and ultracompact \hii regions (Table 1). 
Top: Observed line width. Middle: Thermal line width. Bottom: Non-thermal line width. 
\label{fig-widvsd}}

\figcaption[]{
Map of the radio continuum emission, at 4.9 GHz, from the W3 massive
star forming region (taken from Tieftrunk et al. 1997).
\label{fig-hiigroup}}

\figcaption[]{
Maps of molecular emission from the G34.3+0.15 massive star forming region at 
different angular resolutions. Left: NH$_3$(1,1) emission with $2\pam2$ 
resolution (Heaton et al. 1985). Top right: HCO$^+$(1-0) emission with $6''$ 
resolution (Carral \& Welch 1992). Bottom right: NH$_3$(3,3) emission 
at $1''$ resolution (Heaton et al. 1989). The dashed line corresponds to the lowest 
contour level of the 15 GHz emission from the bright cometary \hii region. 
\label{fig-g34morph}}

\figcaption[]{
Association between ammonia emission from hot molecular gas and radio continuum 
emission from ionized gas, toward three regions of massive star formation. 
Top: G10.47+0.03 (Cesaroni et al. 1998). Middle: G29.96-0.02 (Cesaroni et al. 
1998). Bottom: G61.48+0.09 (G\'omez et al 1995). The radio continuum emission is 
represented in grey scale in the top and middle panels and by dashed lines in 
the lower panel. Ammonia emission is shown by the continuous lines. 
\label{fig-nh3cores}}

\figcaption[]{
Rotational temperature of the ammonia gas toward the G34.3+0.15 massive star 
forming region as a function of distance from the exciting central star
or stars. The line is a least squares linear fit to the data points.
\label{fig-tempvsrad}}

\figcaption[]{
Molecular gas temperature at a fixed distance (0.1 pc) from 
the central exciting stars as a function of luminosity, for the four 
regions with reported temperature profiles with radius (see text).
The line is a least squares linear fit to the data points.
\label{fig-tempvslum}}

\figcaption[]{
Observed line width versus diameter relationship for dense molecular cores.
Filled circles and squares: low-mass cores (Myers 1983); open circles and 
squares: massive cores (Caselli \& Myers 1995); stars: hot cores. 
\label{fig-nh3widvsrad}}

\figcaption[]{Maser spots (H$_2$O: crosses; OH: boxes; CH$_3$OH:
five pointed stars; NH$_3$: filled triangle) toward the G9.62+0.19 region of 
massive star formation, superimposed on contour maps of the hot molecular 
gas (NH$_3(5,5)$ emission: dash line) and ionized gas (radio continuum 
emission: continuous line). 
\label{fig-maser}} 


\end{document}


\newpage
\clearpage

\begin{deluxetable}{lccccrc}
\tablecolumns{7}
\tablewidth{0pt}
\tablecaption{Parameters of compact H II regions \label{tbl-obs1}}
\tablehead{
\colhead{Source} & \colhead{Diam.} & \colhead{$n_e$} & 
\colhead{$\Delta v$} & \colhead{Line} & \colhead{$T_e$} & \colhead{Ref.} \\
\colhead{} & \colhead{(pc)} & \colhead{($10^4$ cm$^{-3}$)} & 
\colhead{(\kms)} &  & \colhead{(K)} & 
\colhead{} 
}
\startdata
DR21-A       & 0.05  &  2.7  & 30.1 & H76$\alpha$ & 7200 & 1 \\
DR21-B       & 0.06  &  3.0  & 30.6 & H76$\alpha$ & 7540 & 1 \\
DR21-C       & 0.07  &  3.0  & 31.0 & H76$\alpha$ & 7010 & 1 \\
DR21-D       & 0.05  &  2.3  & 25.7 & H76$\alpha$ & 7550 & 1 \\
DR21-E       & 0.08  &  0.8  & 43.0 & H76$\alpha$ & 10100 & 1 \\
DR21-F       & 0.10  &  0.5  & 50.0 & H76$\alpha$ & 10480 & 1 \\
G5.89-0.39   & 0.12  &  2.7  & 64.0 & H76$\alpha$ & 8000 & 2 \\
G5.89-0.39:A & 0.044 & 23.7  & 49.0 & H76$\alpha$ & 7200 & 3 \\
G5.89-0.39:B & 0.044 & 23.7  & 56.7 & H76$\alpha$ & 7200 & 3 \\
G10.62-0.38  & 0.048 & 15.6  & 34.3 & H76$\alpha$ & 8000 & 3 \\
G11.94-0.62  & 0.043 &  6.1  & 23.9 & H76$\alpha$ & 7300 & 3 \\
G29.96-0.02  & 0.109 &  4.4  & 29.9 & H76$\alpha$ & 6000 & 3 \\
G30.54+0.0A  & 0.167 &  1.8  & 38.4 & H76$\alpha$ & 6400 & 3 \\
G34.3+0.2    & 0.068 & 13.0  & 55.5 & H76$\alpha$ & 7500 & 4 \\
G35.20-1.74  & 0.039 &  8.6  & 31.8 & H76$\alpha$ & 7100 & 3 \\
G45.07+0.13  & 0.040 & 16.0  & 48.1 & H76$\alpha$ & 8700 & 4 \\
G45.12+0.13  & 0.119 & 11.3  & 47.8 & H76$\alpha$ & 7900 & 3 \\
GM24         & 0.04  &  3.8  & 42.8 & H76$\alpha$ & 10000 & 5 \\
K3-50-A      & 0.09  &  7.9  & 49.0 & H76$\alpha$ & 7900 & 6 \\
K3-50-C1     & 0.06  &  3.8  & 35.0 & H76$\alpha$ & 7100 & 6 \\
NGC6334-A    & 0.20  &  0.31 & 23.9 & H76$\alpha$ & 6900 & 7 \\
NGC6334-F    & 0.02  &  3.8  & 32.0 & H76$\alpha$ & 6400 & 7 \\
W3-A         & 0.40  &  0.84 & 28.4 & H76$\alpha$ & 12900 & 1 \\
W3-B         & 0.186 &  1.7  & 27.5 & H76$\alpha$ & 9000 & 1 \\
W3-C         & 0.082 &  1.96 & 26.7 & H76$\alpha$ & 9100 & 1 \\
W3(OH)       & 0.015 & 13.0  & 42.1 & H76$\alpha$ & 8000 & 8 \\
W49S-K       & 0.22  &  2.2  & 39.9 & H76$\alpha$ & 6800 & 8 \\
W51-d        & 0.20  &  2.0  & 27.0 & H76$\alpha$ & 6600 & 8 \\
W51-e        & 0.33  &  1.0  & 25.4 & H76$\alpha$ & 6400 & 8 \\
G10.6-0.4    & 0.086 &  3.9 & 33.3 & H66$\alpha$ & 7300 & 8 \\
G19.6-0.2A   & 0.053 &  2.4 & 29.5 & H66$\alpha$ & 5400 & 8 \\
G20.1-0.1A   & 0.039 &  2.9 & 29.1 & H66$\alpha$ & 8600 & 8 \\
G34.3+0.2    & 0.033 & 10.0 & 54.7 & H66$\alpha$ & 8000 & 4 \\
G45.07+0.13  & 0.033 & 12.0 & 42.3 & H66$\alpha$ & 7100 & 4 \\
M17-UC1      & 0.0064 & 33. & 47.0 & H66$\alpha$ & 6600 & 9 \\
\tablebreak
N7538-IRS1N  & 0.0026 & 10. & 180.0 & H66$\alpha$ & 15000 & 10 \\
N7538-IRS1S  & 0.0025 & 10. & 180.0 & H66$\alpha$ & 15000 & 10 \\
Sgr-N-K1     & 0.024 & 12.0 & 31.3 & H66$\alpha$ & 8800 &  11,12 \\
Sgr-N-K3     & 0.013 & 28.0 & 37.6 & H66$\alpha$ & 9400 & 11,12 \\
Sgr-N-K4     & 0.04 &   7.7 & 36.0 & H66$\alpha$ & 6800 & 11,12 \\
Sgr-N-K5     & 0.21 &   2.8 & 29.7 & H66$\alpha$ & 5900 & 11,12 \\
Sgr-N-K6     & 0.39 &   2.0 & 47.8 & H66$\alpha$ & 5100 & 11,12 \\
Sgr-N-L      & 0.22 &   3.4 & 31.7 & H66$\alpha$ & 5100 & 11,12 \\
Sgr-N-L13.30 & 0.06 &   3.4 & 20.6 & H66$\alpha$ & 4700 & 11,12 \\
Sgr-N-R      & 0.31 &   1.0 & 31.9 & H66$\alpha$ & 4700 & 11,12 \\
Sgr-A1       & 0.26 &   3.3 & 36.4 & H66$\alpha$ & 5700 & 11,13 \\
Sgr-A2       & 0.06 &   4.4 & 29.7 & H66$\alpha$ & 6700 & 11,13 \\
Sgr-B        & 0.035 & 12.7 & 40.7 & H66$\alpha$ & 7700 & 11,13 \\
Sgr-B9.89    & 0.019 &  6.2 & 50.1 & H66$\alpha$ & 4300 & 11,13 \\
Sgr-B9.96    & 0.022 &  5.5 & 26.5 & H66$\alpha$ & 4200 & 11,13 \\
Sgr-B9.99    & 0.028 &  5.4 & 32.3 & H66$\alpha$ & 6500 & 11,13 \\
Sgr-B10.06   & 0.015 &  8.5 & 35.7 & H66$\alpha$ & 4500 & 11,13 \\
Sgr-B10.10   & 0.020 &  6.5 & 28.9 & H66$\alpha$ & 4400 & 11,13 \\
Sgr-C        & 0.03 &   6.2 & 27.3 & H66$\alpha$ & 6300 & 11,13 \\
Sgr-D        & 0.03 &   8.4 & 34.1 & H66$\alpha$ & 7200 & 11,13 \\
Sgr-E        & 0.06 &   9.4 & 34.8 & H66$\alpha$ & 7000 & 11,13 \\
Sgr-F1       & 0.021 & 26.0 & 61. & H66$\alpha$ & 26400 & 11,13 \\
Sgr-F2       & 0.022 & 22.0 & 75. & H66$\alpha$ & 26400 & 11,13 \\
Sgr-F3       & 0.022 & 43.0 & 63. & H66$\alpha$ & 26400 & 11,13 \\
Sgr-F4       & 0.017 & 26.0 & 59. & H66$\alpha$ & 26400 & 11,13 \\
Sgr-G        & 0.015 & 26.0 & 44.3 & H66$\alpha$ & 14900 & 11,13 \\
Sgr-H        & 0.06 &  13.0 & 45.1 & H66$\alpha$ & 6100 & 11,13 \\
Sgr-I        & 0.08 &   8.3 & 36.6 & H66$\alpha$ & 7300 & 11,13 \\
Sgr-J        & 0.26 &   1.7 & 34.0 & H66$\alpha$ & 5600 & 11,13 \\
Sgr-Y        & 0.06 &   3.0 & 24.0 & H66$\alpha$ & 5500 & 11,13 \\
Sgr-Z        & 0.20 &   1.3 & 24.6 & H66$\alpha$ & 5700 & 11,13 \\
W49-A        & 0.070 &  5.5 & 53.7 & H66$\alpha$ & 10400 & 14 \\
W49-B        & 0.062 &  4.5 & 63.9 & H66$\alpha$ & 17600 & 14 \\
W49-C        & 0.062 &  6.0 & 39.5 & H66$\alpha$ & 11500 & 14 \\
\tablebreak
W49-D        & 0.070 &  5.6 & 38.2 & H66$\alpha$ & 7800 & 14 \\
W49-F        & 0.058 &  5.5 & 26.2 & H66$\alpha$ & 13400 & 14 \\
W49-G1       & 0.112 &  3.4 & 43.3 & H66$\alpha$ & 10500 & 14 \\
W49-G2       & 0.122 &  3.3 & 48.6 & H66$\alpha$ & 15400 & 14 \\
W49-G3       & 0.160 &  3.1 & 37.0 & H66$\alpha$ & 7200 & 14 \\
W49-G4       & 0.140 &  3.1 & 32.8 & H66$\alpha$ & 8100 & 14 \\
W49-G5       & 0.152 &  3.0 & 36.6 & H66$\alpha$ & 8900 & 14 \\
W49-I        & 0.182 &  1.1 & 35.6 & H66$\alpha$ & 8000 & 14 \\
W49-J        & 0.168 &  1.7 & 27.3 & H66$\alpha$ & 11700 & 14 \\
W49-L        & 0.96 &   0.4 & 31.0 & H66$\alpha$ & 10000 & 14 \\
W49-M        & 0.160 &  1.2 & 37.1 & H66$\alpha$ & 6700 & 14 \\
W49-N        & 0.070 &  2.3 & 35.3 & H66$\alpha$ & 14900 & 14 \\
W49-O        & 0.082 &  5.0 & 37.3 & H66$\alpha$ & 5800 & 14 \\
W49-S        & 0.174 &  1.9 & 39.0 & H66$\alpha$ & 8000 & 14 \\
W3(OH)       & 0.015 & 13.0 & 38.3 & H66$\alpha$ & 8000 & 8 \\
\enddata
\tablenotetext{ }{References: 
(1) Roelfsema 1987;
(2) Zjilstra et al. 1990; 
(3) Wood \& Churchwell 1989a;
(4) Garay, Rodr\'\i guez \& van Gorkom 1986;
(5) Roth et al. 1988;
(6) De Pree et al. 1994;
(7) De Pree et al. 1995a;
(8) Garay, Reid, \& Moran 1985;
(9) Johnson, De Pree, \& Goss 1998;
(10) Gaume et al. 1995b;
(11) Gaume et al. 1995a;
(12) De Pree et al. 1995c;
(13) De Pree et al. 1996;
(14) De Pree, Mehringer \& Goss 1997.}
\end{deluxetable}

\vfill\eject

\begin{deluxetable}{lllccccc}
\tablecolumns{8}
\tablewidth{0pt}
\tablecaption{Hot ammonia cores~$^{\rm a}$ \label{tbl-hotcores}}
\tablehead{
\colhead{Source} & \colhead{Dist.} & \colhead{Radius} & \colhead{$\Delta v$} 
& \colhead{N(NH$_3$)} & \colhead{n(H$_2$)} & \colhead{Mass~$^{\rm b}$} & 
\colhead{Ref.} \\
\colhead{} & \colhead{(kpc)} & \colhead{(pc)} & \colhead{(\kms)} & 
\colhead{(cm$^{-2}$)} & \colhead{(cm$^{-3}$)} & \colhead{(M$_{\odot}$)} & 
\colhead{} 
}
\startdata
G5.89-0.39.... & 4.0  & 0.2    & 8.0 &  $9\times10^{15}$  & $6\times10^3$   & 30$^m$  & 1 \\
G9.62+0.19.... & 5.7  & 0.066  & 5.1 & $1.6\times10^{18}$ & $1.2\times10^7$ & 164$^v$ & 2 \\
               &      & 0.040  & 4.3 &                    &                 &         & 2 \\
G10.47+0.03... & 5.8  & 0.063  & 5.4 & $1.4\times10^{19}$ & $1.0\times10^7$ & 218$^v$ & 2 \\
               &      & 0.048  & 5.7 &                    &                 &         & 2 \\
G10.47+0.03c   & 5.8  & 0.042  & 8.1 & $3.2\times10^{18}$ & $1.2\times10^7$ & 210$^m$ & 3 \\
               &      & 0.042  & 9.0 &                    &                 &         & 3 \\
G19.61-0.23... & 3.5  & 0.026  & 9.5 & $2.4\times10^{17}$ & $9\times10^6$   &  25$^m$ & 4 \\
G29.96-0.02... & 7.4  & 0.072  & 7.6 & $1.7\times10^{18}$ & $1.3\times10^7$ & 138$^v$ & 2 \\
               &      & 0.038  & 4.3 &                    &                 &         & 2 \\
G31.41+0.31... & 7.9  & 0.073  & 5.1 & $8.2\times10^{18}$ & $8.6\times10^6$ & 285$^v$ & 2 \\
               &      & 0.054  & 5.8 &                    &                 &         & 2 \\
G32.80+0.19S   & 13.  & 0.079  & 7.1 &  $3\times10^{16}$  & $5\times10^4$   &  18$^m$ & 5 \\
G32.80+0.19N   & 13.  & 0.10   & 4.8 &  $5\times10^{15}$  &  $7\times10^3$  &  10$^m$ & 5 \\
G32.80+0.19M   & 13.  & 0.063  & 6.5 &                    &                 &         & 5 \\
G34.26+0.15... & 3.8  & 0.015  & 7.5 &  $7\times10^{18}$  & $7.0\times10^7$ &         & 6 \\
               &      & 0.027 &$\sim$5& $7\times10^{18}$  & $4.2\times10^7$ & 170$^v$ & 7 \\
G61.48W        & 5.4  & 0.039  & 4.2 &  $1\times10^{16}$  & $4\times10^4$   &   2$^m$ & 5 \\
G61.48E        & 5.4  & 0.020  & 3.5 &  $6\times10^{15}$  & $3\times10^4$   & 0.6$^m$ & 5 \\
SgrB2-M        & 8.0  & 0.1    &15.0 &  $9\times10^{17}$  & $2\times10^5$   & 50$^v$  & 8 \\ 
SgrB2-N        & 8.0  & 0.1    &19.5 &  $1\times10^{20}$  & $1\times10^7$   & 5000$^v$ & 8 \\ 
W3(H2O)        & 2.2  & 0.006  & 4.2 &  $1\times10^{17}$  & $4\times10^8$   & 18$^m$  & 9 \\ 
W3IRS5         & 2.3  & 0.03   & 5.6 & $1.5\times10^{15}$ & 1-4$\times10^7$ & 60-240$^v$ & 10 \\
\enddata
\tablenotetext{a}{Parameters derived from observations of ammonia lines made with 
high angular resolution (VLA).}
\tablenotetext{b}{In this column superscript $m$ denotes
molecular mass derived as in eqn.(12) and superscript $v$ denotes virial mass.}
\tablenotetext{ }{References: 
(1) G\'omez et al. 1991;
(2) Cesaroni et al. 1994a;
(3) Garay, Moran \& Rodr\'\i guez 1993;
(4) Garay et al. 1998;
(5) G\'omez, Garay \& Lizano 1995;
(6) Garay \& Rodr\'\i guez 1990;
(7) Heaton, Little, \& Bishop 1989;
(8) Vogel, Genzel, \& Palmer 1987;
(9) Wilson, Gaume, \& Johnston 1993;
(10) Tieftrunk, Gaume \& Wilson 1998b.
}
\end{deluxetable}

\begin{deluxetable}{lccrccl}
\tablecolumns{7}
\tablewidth{0pt}
\tablecaption{Young massive objects associated with disks and outflows \label{tbl-ymo}}
\tablehead{
\colhead{Source} & \colhead{Luminosity} & \colhead{$M_{disk}$} & 
\colhead{$R_{disk}$} & \colhead{Molec.} & \colhead{Jet?} & \colhead{Ref.} \\
\colhead{} & \colhead{(\Lsun)} & \colhead{(\Msun)} & 
\colhead{(AU)} & \colhead{outflow?}  & & 
\colhead{} 
}
\startdata
W3(H$_2$O)      & $(1-10)\times10^3$& 10-20 & $<500$ & y & y & 1,2 \\ 
GL 490          & $2\times10^3$     &  10  & 8500  & y & n & 3,4,5 \\ 
AFGL 5142       & $4\times10^3$     & 150  & 6000 & y & n & 6 \\
G192.16-3.82    & $3\times10^3$     &  15  & 1000  & y & y & 7,8 \\
HH 80-81        & $1.7\times10^4$   &  30  & $<4000$ & y & y & 9,10,11 \\
IRAS 20126+4104 & $1.3\times10^4$   &  10  &  850 & y & y & 12,13,14 \\   
Cep A HW2       & $1\times10^4$     & 200  &  750 & y & y & 15,16 \\
IRAS 23385+6053 & $1.6\times10^4$   & 370  & 10000 & y & n & 17 \\
\enddata
\tablenotetext{ }{References: 
(1) Wilner et al. 1995;
(2) Reid et al. 1995;
(3) Nakamura et al. 1991;
(4) Snell et al. 1984;
(5) Chini, Henning, \& Pfau 1991;
(6) Hunter et al. 1999;
(7) Shepherd \& Kurtz 1999;
(8) Shepherd et al. 1998;
(9) Yamashita et al. 1991;
(10) Yamashita et al. 1989;
(11) Mart\'\i , J., Rodr\'\i guez, L.F., \& Reipurth, B. 1993;
(12) Cesaroni et al. 1997;
(13) Hofner et al. 1999;
(14) Zhang et al. 1998b;
(15) G\'omez et al. 1999;
(16) Torrelles et al. 1996;
(17) Molinari et al. 1998.} 
\end{deluxetable}